\documentclass[aps,prl,showpacs,twocolumn,groupedaddress]{revtex4}  
\usepackage{graphicx}   
\usepackage{dcolumn}    
\usepackage{bm}         
\usepackage{epsfig}     

\def \ttbar {\ensuremath{ t\bar{t}                  }}

\def \ifb    {\ensuremath{ \rm fb^{-1}                          }}

\def \pt     {\ensuremath{ p_T                                  }}

\def \mtmin {\ensuremath{M_T^{\rm min}}}
\def \etmisssc {\mbox{\ensuremath{E\kern-0.6em\slash_T^{\rm\kern+0.1em Sc}}}}
\def \etmiss {\mbox{\ensuremath{E\kern-0.6em\slash_T}}}

\newcommand{\Eslash}{\mbox{$E \kern-0.6em\slash$                }}

\begin{document}


\title{
Search for Higgs boson production in dilepton and missing energy
final states with ~5.4 $\bm{\mathrm{fb^{-1}}}$ of $\bm{p\bar{p}}$ 
collisions at $\bm{\sqrt s =1.96}$ TeV}

%
\author{V.M.~Abazov$^{37}$}
\author{B.~Abbott$^{75}$}
\author{M.~Abolins$^{64}$}
\author{B.S.~Acharya$^{30}$}
\author{M.~Adams$^{50}$}
\author{T.~Adams$^{48}$}
\author{E.~Aguilo$^{6}$}
\author{G.D.~Alexeev$^{37}$}
\author{G.~Alkhazov$^{41}$}
\author{A.~Alton$^{64,a}$}
% ...

\date{\today}

\begin{abstract}
A search for the standard model Higgs boson is presented
using events with two charged leptons and large missing transverse
energy selected from 5.4~\ifb\ of integrated luminosity in $p \overline{p}$ collisions 
at $\sqrt{s}=1.96$ TeV collected with the D0 detector at the 
Fermilab Tevatron collider. No significant
excess of events above background predictions is found, and 
observed (expected) upper limits at 95\% confidence level
on the rate of Higgs boson production are derived that are a factor of 1.55 (1.36) above the 
predicted standard model cross section at $m_H$=165~GeV.
\end{abstract}


\maketitle

The Higgs mechanism, introduced in the standard model (SM) to explain electroweak symmetry breaking, predicts a massive scalar (Higgs) boson,
which has yet to be observed.
Direct searches at the CERN LEP $e^+e^-$ collider yielded
a lower limit of 114.4~GeV for the SM Higgs boson mass at 95\% confidence level (C.L.)~\cite{lephiggs}.
Indirect constraints obtained from fits to precision electroweak data, when combined with direct searches at LEP,
give an upper bound of $186$~GeV at 95\% C.L.~\cite{lepewwg_summer08}.
For a Higgs boson mass ($m_H$) close to 165~GeV the product of the 
SM Higgs boson production cross section and the decay branching ratio into 
two $W$ bosons is maximal~\cite{turcot} and motivates the analysis strategy.

In this Letter we present a search for Higgs
bosons in final states containing two charged leptons and
missing transverse energy (\etmiss) using data collected
with the D0 detector~\cite{d0det} and corresponding to
an integrated luminosity of 5.4~\ifb\ of $p\bar{p}$ 
collisions at $\sqrt s =1.96$ TeV. We consider final states
containing either an electron and a positron ($e^+e^-$), an electron or a positron and a
muon ($e^{\pm} \mu^{\mp}$), or two muons  ($\mu^+\mu^-$). 
Final states with tau leptons decaying to $e$ or $\mu$ or where hadronic 
tau decays are misidentified as electrons will also contribute to our search.

Previous searches in this channel have been performed at the Tevatron by the CDF and the D0 collaborations~\cite{cdf_hww_3fb,hohlfeld}.
This search represents an almost twentyfold increase in the D0 data set 
and considers additional Higgs boson production modes leading to the 
dilepton and {\etmiss} signature. 
In addition, the lepton acceptance is improved and the separation of background 
and signal processes now utilizes an artificial neural network (NN) event classification technique. 
The main Higgs boson production modes are via gluon fusion and vector boson 
fusion. For these production modes, this analysis considers only the Higgs boson decay 
$H \rightarrow W W^{(*)} \rightarrow \ell \ell'\nu\nu'~(\ell, \ell' = e, \mu, \tau)$.
Also considered is Higgs boson production in association with a $W$ or $Z$
boson, where Higgs boson decays to $W/Z$ bosons 
and leptons yield a dilepton plus {\etmiss} signature.
The overlap with events considered in the analysis of $WH\rightarrow Wb\overline{b}$ and $ZH\rightarrow Zb\overline{b}$ 
final states~\cite{VH} is negligible.  The CDF collaboration is also reporting an updated search in this channel~\cite{cdfhww}.

The main background processes for this analysis are pair production of 
heavy gauge bosons, $W(\rm{ +jets/}\gamma)$ and $Z/\gamma^*(\rm{ +jets/}\gamma)$ production,
$t{\bar t}$ production and multijet production in which jets
 are misidentified as leptons. To model the $W(\rm{ +jets/}\gamma)$ and $Z/\gamma^*(\rm{ +jets/}\gamma)$ backgrounds we use the
\textsc{alpgen} event generator \cite{alpgen}. 
The signal and remaining SM background processes are simulated with
\textsc{pythia}~\cite{pythia} and all Monte Carlo (MC) samples are generated using CTEQ6L1~\cite{cteq6}
parton distribution functions (PDFs). 
In all cases, event generation is followed by a detailed {\sc geant}-based~\cite{geant} simulation of the D0 detector.

The background MC samples for inclusive $W$ and $Z/\gamma^*$ production
are normalized to next to next to leading order (NNLO) cross section predictions~\cite{hamberg} calculated
using MRST 2004 NNLO PDFs~\cite{MRST}.
The rate of \ttbar\ production is normalized to a NNLO calculation~\cite{moch} and diboson rates 
($WW$, $WZ$, and $ZZ$) are normalized to next to leading order (NLO) cross sections~\cite{ellis}. 
The signal cross sections are calculated at NNLO~\cite{higgs_xsec}
(at NLO in the case of the vector boson fusion process).
The branching fractions for the Higgs boson decay 
are determined using {\sc hdecay}~\cite{hdecay}.

The simulated $Z$ boson transverse momentum (\pt) distribution is modified to match 
the spectrum measured in data~\cite{zpt}.
In order to simulate the $W$ boson \pt\ distribution, the measured $Z$
boson \pt\ spectrum is multiplied by the ratio of $W$ to $Z$ boson \pt\ distributions at 
NLO~\cite{melnikov}.
To improve the modeling of $WW$ background, 
the \pt\  of the diboson system is modified to
match that obtained using the \textsc{mc@nlo} generator~\cite{mcatnlo}, and the
distribution of the opening angle of the two leptons is modified to take 
into account the contribution from gluon-gluon initiated processes~\cite{binoth}.
The Higgs boson transverse momentum distribution in the \textsc{pythia}-generated gluon fusion sample is modified to match the distribution obtained using \textsc{sherpa}~\cite{sherpa}.

The background due to multijet production, in which 
jets are misidentified as leptons, is determined from data. For this purpose, a sample of like-charged dilepton 
events is used in the $\mu^+\mu^-$ channel, corrected for like-charge contributions from non-multijet processes.
The $e^+e^-$ and $e^{\pm} \mu^{\mp}$ channels use a sample of events with inverted lepton quality requirements,
scaled to match the yield and kinematics determined in the like-charge data.

This search is based on a sample of dilepton event candidates collected using a mixture
of single and dilepton triggers which achieve close to 100\% signal efficiency. 
The identification of electron and muon candidates is based on the criteria described
in the previous search~\cite{hohlfeld}. In addition to the track isolation criterion, 
a constraint on the scalar sum of charged particles transverse momentum ($p_T$) in a cone of 
radius ${\cal R} = \sqrt{(\Delta\phi)^2+(\Delta\eta)^2} = 0.5$~\cite{d0coord} around the muon 
track, an isolation requirement in the calorimeter is applied. This is a requirement on the transverse 
energy deposited in an annulus $0.1< {\cal R} <0.4$ around the muon track.
In the $e^{\pm} \mu^{\mp}$ channel, each of these isolation parameters divided by the
muon \pt\ is required to be $<0.15$, whereas in the $\mu^+\mu^-$ channel the ratio of the sum of these two quantities
divided by the muon \pt\ is 
required to be $<0.4 (0.5)$ for the highest (next-to-highest)~\pt\ lepton $\ell_1$ ($\ell_2$). 
In the $\mu^+\mu^-$ channel, the product of the
isolation ratios for both muons is required to be $<0.06$. 

Electrons are required to have $|\eta|<2.5$ ($<2.0$ in the $e^+e^-$ channel),
and muons $|\eta|<2.0$.
Both leptons are required to originate from the same interaction
vertex and to have opposite charges. Electrons must have
$p_T^{e} >15$~GeV, and muons $p_T^{\mu}>10$~GeV.
In the $\mu^+\mu^-$ channel one of the two muons is required to 
have $p_T^{\mu} >20$~GeV.  In addition, the dilepton invariant mass 
is required to exceed 15~GeV. Jets are
reconstructed in the calorimeter using an iterative midpoint cone algorithm \cite{jets}
with a radius ${\cal R}=0.5$ and are required to have $p_T^{\rm jet} >15$~GeV 
and $|\eta| < 2.4$. No jet-based event selection is applied, since the number 
of jets in the event is used in the NN to help discriminate signal from background.
In the $\mu^+\mu^-$ channel, both muons must be 
separated from any jet by ${\cal R} > 0.1$. This
stage of the analysis is referred to as ``preselection".

\begin{figure*}[htb]
  \begin{center}
  \begin{tabular}{lcr}
  \includegraphics[width=0.675\columnwidth]{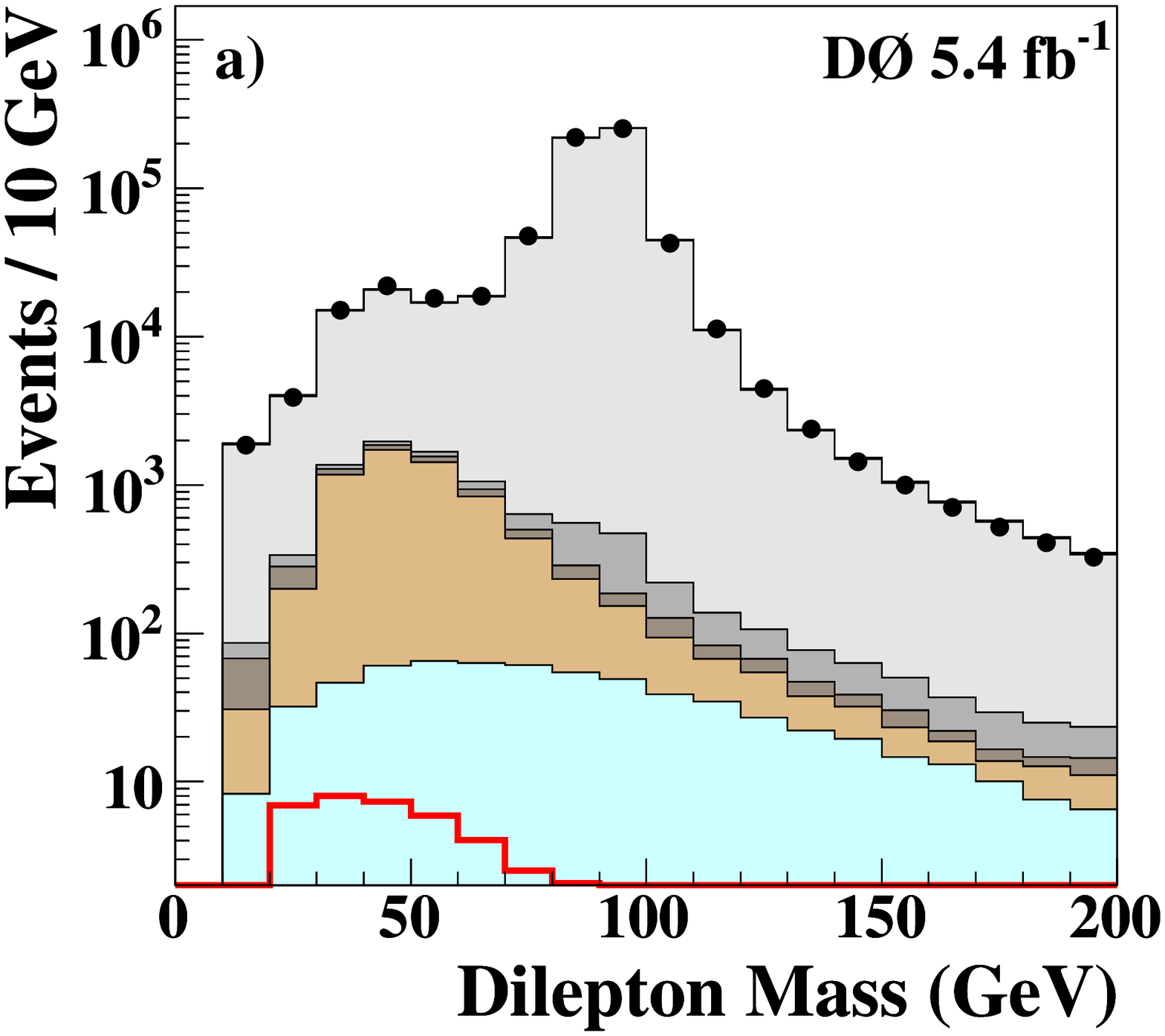} &
  \includegraphics[width=0.675\columnwidth]{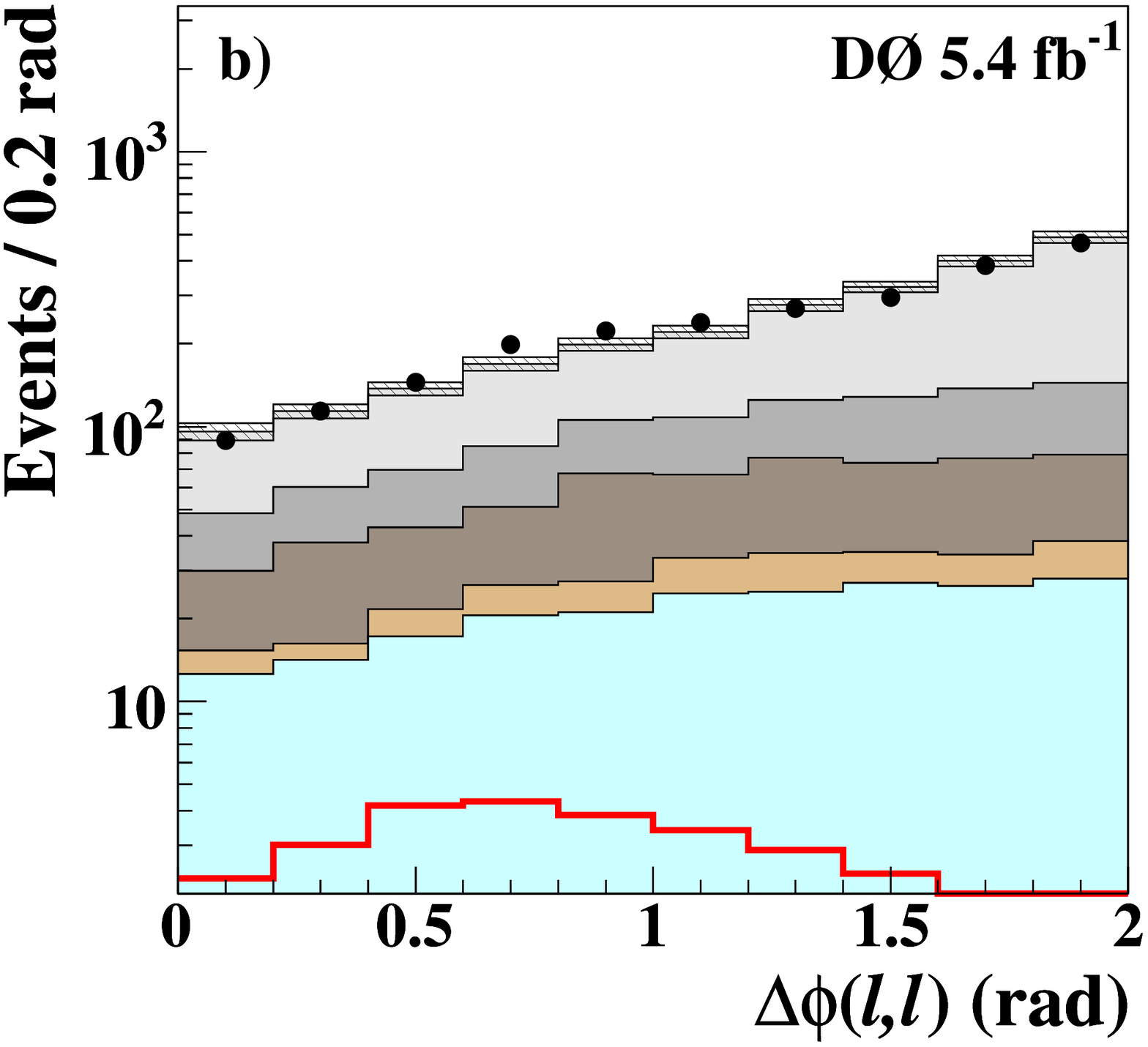} &
  \includegraphics[width=0.675\columnwidth]{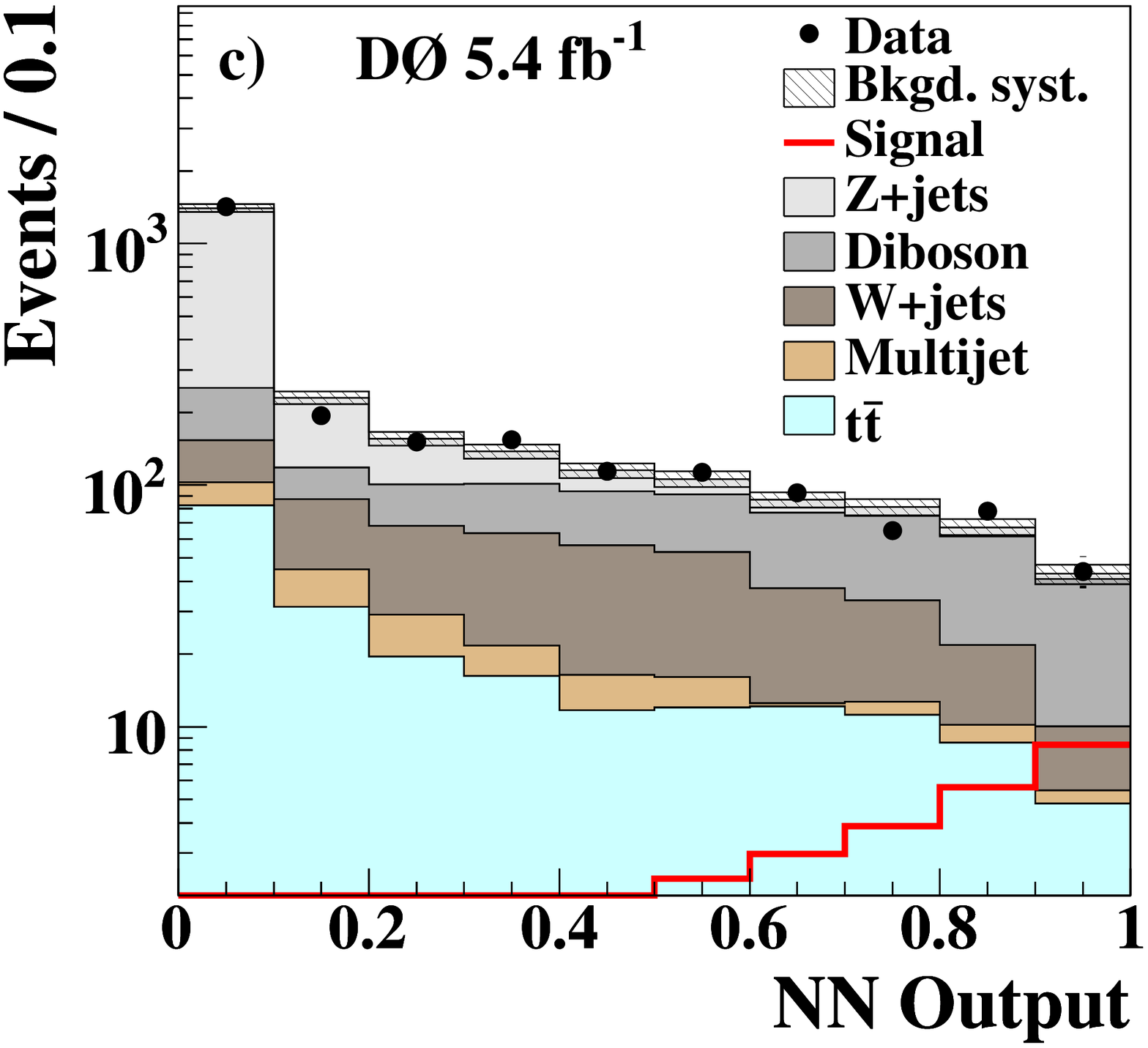} \\
  \end{tabular}
  \end{center}
  \caption{(color online) (a) The dilepton invariant mass after preselection;
  (b) the $\Delta\phi(\ell,\ell)$ angle after final selection;  
  and (c) the neural network output after final selection.
  The signal is shown for $m_H$=165~GeV.
  The systematic uncertainty is shown after fitting (see text for details).
  \label{fig:datamc}}
\end{figure*}

After preselection, the background is dominated by $Z/\gamma^*$ production.
This background is suppressed by requiring $\etmiss>20$~GeV
($>25$~GeV in the $\mu^+\mu^-$ channel).
Events are also removed if the \etmiss\ was likely produced by a
mismeasurement of jet energies by requiring for the scaled $\etmiss$~\cite{hohlfeld}, $\etmisssc>6$ in the 
$e^+e^-$ and $e^{\pm}\mu^{\mp}$ channels.
The minimum transverse mass, \mtmin  (defined as the smaller of the transverse masses
$M_T$~\cite{mtdef} calculated from the \etmiss\ and either of the two leptons), is required
to be $>20$~GeV ($>30$~GeV in the $e^+e^-$ channel) to suppress
backgrounds where \etmiss\ originates from mismeasured lepton energy.
Finally, events are rejected by requiring for the azimuthal opening angle 
between the two leptons $\Delta \phi(\ell,\ell)<2.0$~rad, because leptons from
background processes tend to be back-to-back in the transverse plane, in
contrast with those from a Higgs boson decay which, owing to its zero spin,
tend to move in the same direction. This stage of the analysis is referred to
as ``final selection".

The dilepton invariant mass distribution after preselection for the combination 
of the three channels is shown in Fig.~\ref{fig:datamc}(a). 
The $\Delta\phi(\ell,\ell)$ distribution after final selection
is shown in Fig.~\ref{fig:datamc}(b).
The contributions from the different
background processes in each of the three channels are compared with
the numbers of events observed in data after preselection and after final 
selection in Table~\ref{tab:yields}. The total systematic uncertainty (described below and in the supplemental material) after fitting is shown with correlations appropriately incorporated.

\begin{table*}[hbt]
\begin{center}
\caption{\label{tab:yields} Expected and observed event yields in
each channel after preselection and at the final selection. The systematic uncertainty after fitting is shown for all samples at final selection.
}
\begin{tabular}{c|r@{$\, \,$}lr@{$\,\pm \,$}l|r@{$\, \,$}lr@{$\,\pm \,$}l|r@{$\, \,$}lr@{$\,\pm \,$}l}
\hline \hline
                & \multicolumn{4}{c|}{$e^{\pm}\mu^{\mp}$}&
                \multicolumn{4}{c|}{$e^+e^-$}          &
                \multicolumn{4}{c}{$\mu^+\mu^-$} \\
                & \multicolumn{2}{c}{Preselection} & \multicolumn{2}{c|}{ Final selection}& \multicolumn{2}{c}{Preselection}    & \multicolumn{2}{c|}{Final selection}          & \multicolumn{2}{c}{Preselection} & \multicolumn{2}{c}{Final selection} \\
\hline
$Z/\gamma^*\to e^+e^-$       & \multicolumn{2}{c}{120} & \multicolumn{2}{c|}{$<0.1$} & \multicolumn{2}{c}{274886}& 158 & 13 &     \multicolumn{2}{c}{$-$}         &   \multicolumn{2}{c}{$-$}  \\
$Z/\gamma^*\to \mu^+\mu^-$   & \multicolumn{2}{c}{89}     & 4.3&0.3    &     \multicolumn{2}{c}{$-$}            &    \multicolumn{2}{c|}{$-$}            & \multicolumn{2}{c}{373582}& 1247& 37\\
$Z/\gamma^*\to \tau^+\tau^-$ & \multicolumn{2}{c}{3871}       & 7.1&0.5    & \multicolumn{2}{c}{1441}       & 0.7 &0.1 & \multicolumn{2}{c}{2659}   & 12.0 & 0.7\\
$\ttbar$        &  \multicolumn{2}{c}{312}    & 93.8 & 8.3   & \multicolumn{2}{c}{159}     & 47.0&4.4     & \multicolumn{2}{c}{184}    & 74.6 & 6.8 \\
$W+{\rm jets}/\gamma$        &  \multicolumn{2}{c}{267}   & 112 &9  & \multicolumn{2}{c}{308}    & 122&11   & \multicolumn{2}{c}{236} & 91.5 & 6.5 \\
$WW$            & \multicolumn{2}{c}{455}    & 165&6  & \multicolumn{2}{c}{202}    & 73.9 &6.4    & \multicolumn{2}{c}{272} & 107 & 9 \\
$WZ$            & \multicolumn{2}{c}{23.6} & 7.6&0.2 & \multicolumn{2}{c}{137} & 11.5 &1.0 & \multicolumn{2}{c}{171}    & 21.5&2.0 \\
$ZZ$            & \multicolumn{2}{c}{5.4}  & 0.6&0.1 & \multicolumn{2}{c}{117} & 9.3 &0.9  & \multicolumn{2}{c}{147} & 18.0&1.8 \\
Multijet       & \multicolumn{2}{c}{430}  & 6.4&2.5    & \multicolumn{2}{c}{1370}       & 1.0&0.1    & \multicolumn{2}{c}{408}      & 53.8&10.3 \\
\hline Signal
($m_H=165$~GeV) & \multicolumn{2}{c}{18.8} & 13.5 & 1.5 & \multicolumn{2}{c}{11.2}    & 7.2 &0.8    & \multicolumn{2}{c}{12.7}   & 9.0 & 1.0 \\
\hline
Total background& \multicolumn{2}{c}{5573}   & 397 & 14 & \multicolumn{2}{c}{278620}    & 423 &19  & \multicolumn{2}{c}{377659}   & 1625 & 41 \\
\hline\hline
Data            &  \multicolumn{2}{c}{5566}        &  \multicolumn{2}{c|}{390}        &  \multicolumn{2}{c}{278277}          & \multicolumn{2}{c|}{421}           &  \multicolumn{2}{c}{384083}        & \multicolumn{2}{c}{1613} \\
\hline\hline
\end{tabular}
\end{center}
\end{table*}

To improve the separation between signal and background, an optimized NN is used in each of the three channels.
Several well-modeled discriminant variables are used as inputs
to the NN: the transverse momenta of the leptons, a variable indicating
the quality of the leptons' identification, the transverse momentum and invariant 
mass of the dilepton system, \mtmin, \etmiss, $\etmisssc$, $\Delta \phi(\ell,\ell)$, 
$\Delta \phi(\ell_1,\etmiss)$, $\Delta \phi(\ell_2,\etmiss)$, 
the number of identified jets, and the scalar sum of the transverse momenta 
of the jets.
In each channel, separate NNs are trained for 18 test values of $m_H$ from 115 to 200~GeV
in steps of 5~GeV. The combined distribution of the NN output for $m_H=165$ GeV 
from all three channels is shown in Fig.~\ref{fig:datamc}(c).

\begin{figure*}[htb]
  \begin{center}
  \begin{tabular}{lr}
  \includegraphics[width=0.675\columnwidth]{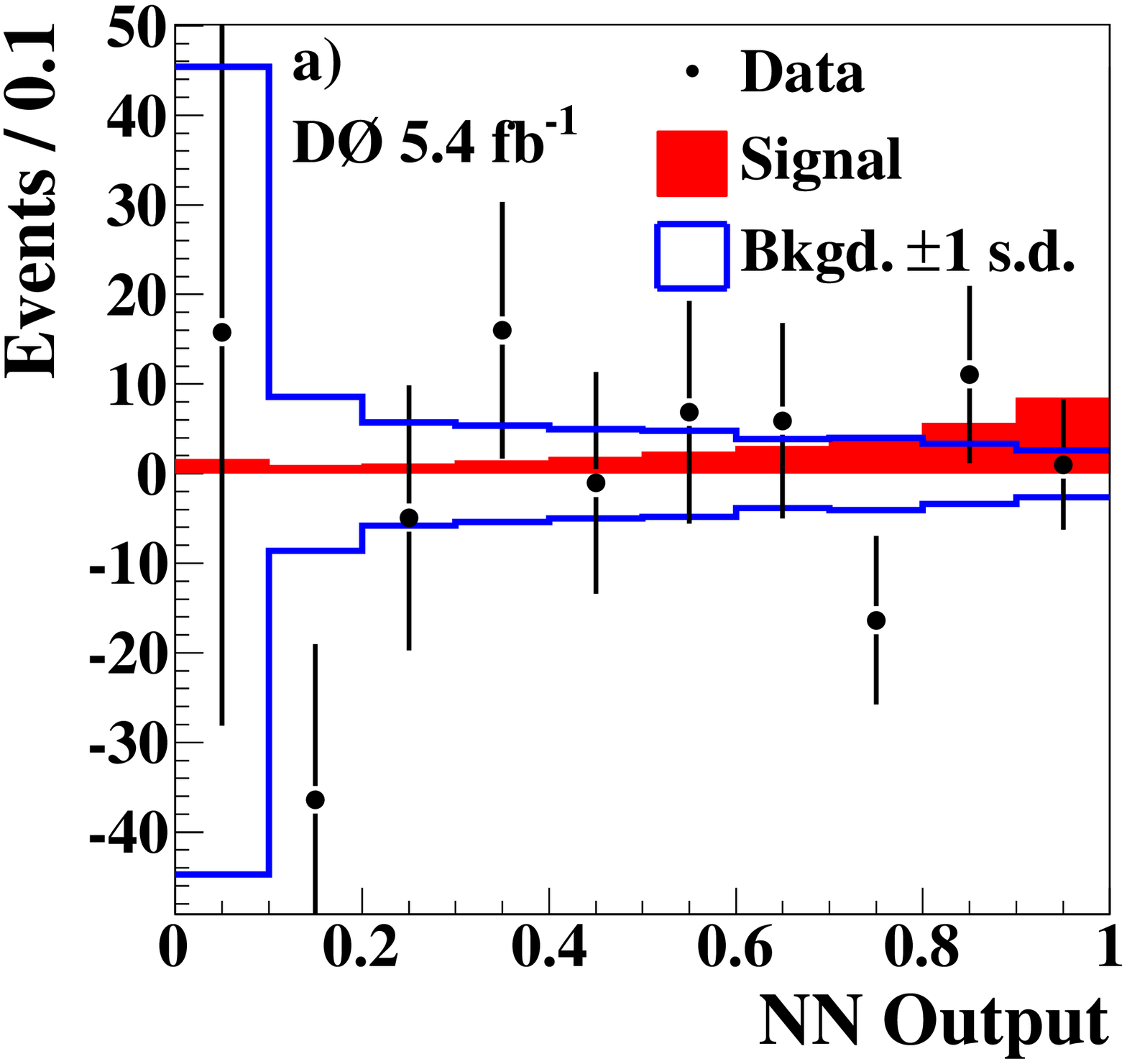} &
  \includegraphics[width=0.675\columnwidth]{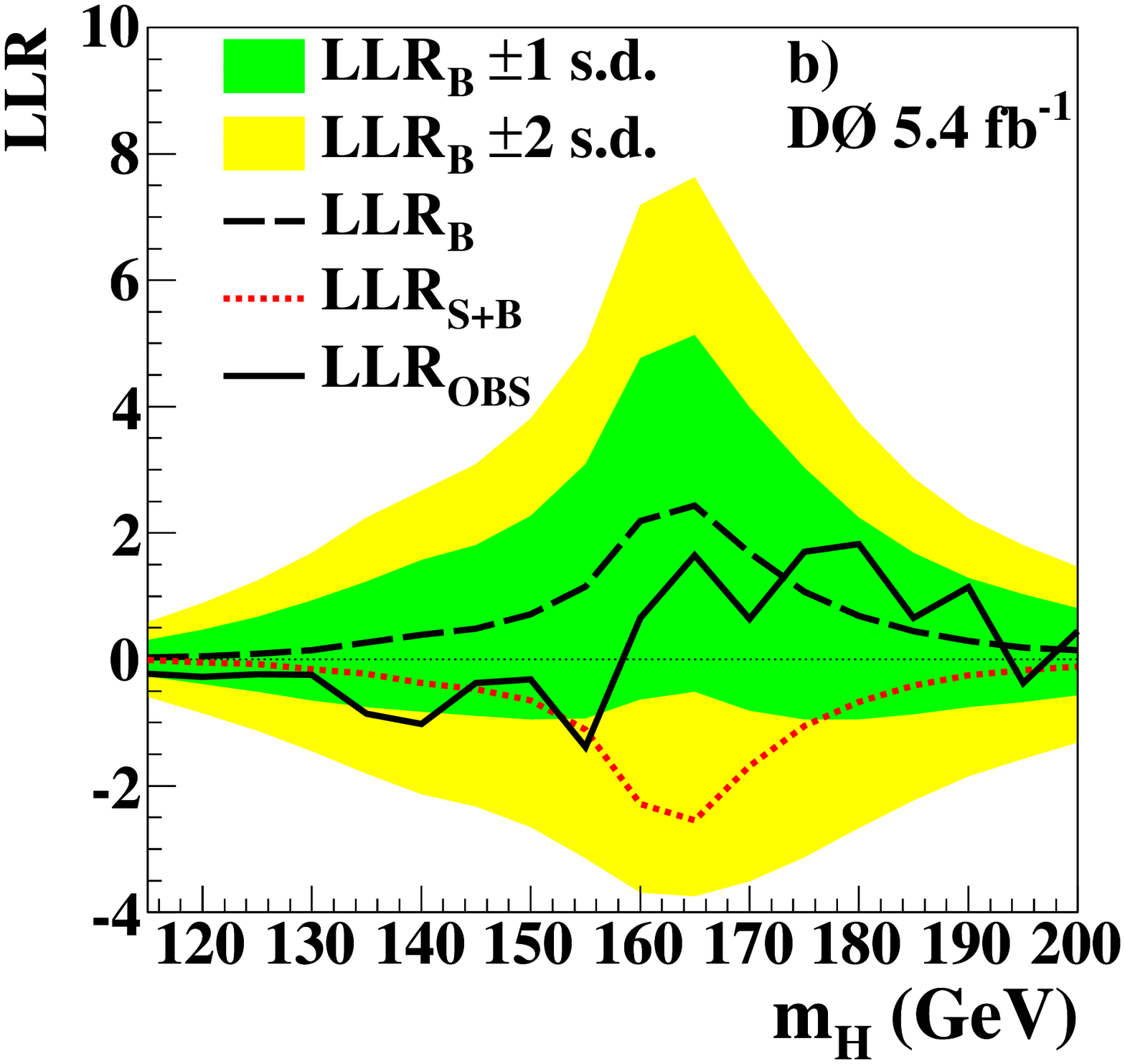}
  \includegraphics[width=0.675\columnwidth]{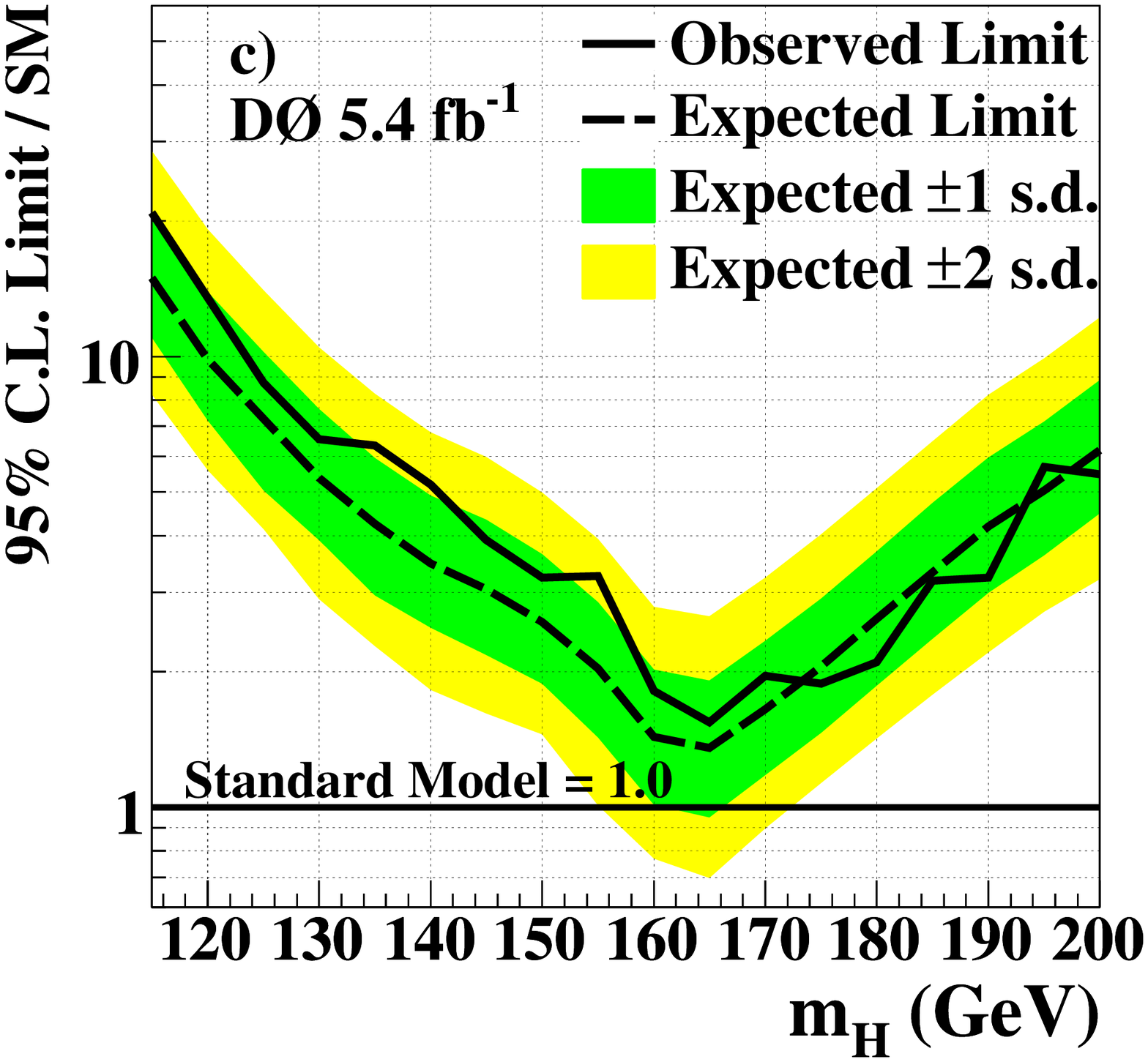}
  \end{tabular}
  \end{center}
  \caption{\label{fig:results} (color online) (a) Data after subtracting the fitted background (points)
    and SM signal expectation (filled histogram) as a function of the NN output
    for $m_H$=165~GeV. Also shown is the $\pm1$ standard deviation (s.d.) band
    on the total background after fitting.
    (b) Observed LLR (solid line), expected LLR for
    background-only hypothesis (dashed line), and signal + background hypothesis (dotted line).
    (c) Upper limit on Higgs boson production cross section at 95\% C.L. expressed as a ratio to the SM cross section. The one and two s.d.~bands around the curve corresponding
    to the background-only hypothesis are also shown.
    }
\end{figure*}

The estimates for the expected number of background and signal events depend on
numerous factors, each introducing a source of systematic uncertainty. Two
types of systematic uncertainties have been considered: those affecting the absolute predicted event yield and those which also affect the shape of the
NN output distribution. The most significant systematic uncertainties
affecting the normalization of the NN output (quoted as a percentage of the yield 
per signal or background process) are:
lepton reconstruction efficiencies (3\%--6\%),
lepton momentum calibration (1\%--3\%), theoretical cross section (including PDF, factorization and renormalization scale uncertainties: 7\% for diboson, 10\% for \ttbar\,7\% for $W/Z$(+jets), 11\% for Higgs signal), modeling of
multijet background (2\%--15), and integrated luminosity (6.1\%). 
The most important sources affecting the NN output shape are:
jet reconstruction efficiency (1\%--3\%), jet energy scale
calibration (1\%--5\%), jet energy resolution (2\%), and
modeling of $\pt(WW)$, $\pt(H)$, and $\pt(Z)$ (1\%--5)\%.
The systematic uncertainty on the modeling of $\pt(WW)$ and $\pt(H)$ has been determined by
comparing the \pt\ distributions of \textsc{pythia},
\textsc{sherpa}, and \textsc{mc@nlo}, and the uncertainty on $\pt(Z)$ from a comparison of the shape of the NN distribution
between data and MC predictions in a $Z/\gamma^*$ enriched control sample. The \textsc{sherpa} and
\textsc{mc@nlo} predictions agree well with each other and generate harder \pt\
spectra than \textsc{pythia}~\cite{Gleisberg:2005qq}.
The uncertainty on $\Delta \phi(\ell,\ell)$ for the $WW$ background 
is taken as 30\% of the correction to the \textsc{pythia}
angular distribution as estimated in Ref.~\cite{binoth}, leading to a relative uncertainty at the subpercent level. 
Appropriate correlations of systematic uncertainties between different channels,
between different backgrounds, and between backgrounds and signal are included.

\begin{table*}[ht]
\caption{\label{tab:alllimitbay} Expected and observed upper limits
at 95\% C.L. for Higgs boson production cross section
expressed as a ratio to the cross section predicted by the SM for a range of test Higgs boson masses.}
\begin{ruledtabular}
\begin{tabular}{ccccccccccccccccccc}
$m_H$ (GeV) & 115& 120& 125& 130& 135& 140& 145& 150& 155& 160& 165& 170& 175& 180& 185& 190& 195& 200 \\
\hline
Limit (exp.) & 14.9    & 9.74    & 7.20    & 5.40   & 4.23   & 3.48   & 3.07   & 2.58   & 2.02   & 1.43   & 1.36   & 1.65   & 2.06   & 2.59   & 3.28   & 4.20   & 5.08   & 6.23   \\
Limit (obs.) & 20.8    & 13.6    & 8.81    & 6.63    & 6.41    & 5.21  &  3.94  &  3.29  &  3.25  &  1.82  &  1.55  &  1.96  &  1.89  &  2.11  &  3.17  &  3.27  &  5.77  &  5.53  \\
\end{tabular}
\end{ruledtabular}
\end{table*}

After all selections, no significant excess of signal-like events is
observed for any test value of $m_H$.
Thus the NN output distributions are used to set upper limits on the Higgs boson production cross section, 
assuming the SM-predicted ratio of production cross sections and Higgs decay branching ratios. Upper limits are set
using the three search channels combined using a modified
frequentist method with a log-likelihood ratio (LLR)
test statistic \cite{bib:limits}. To minimize the degrading effects of
systematics on the search sensitivity, the signal and different background sources
contributions are fitted to the data observations by maximizing a
likelihood function over the systematic uncertainties for both the background-only and signal+background hypotheses~\cite{bib:sys}.
Fig.~\ref{fig:results}(a) shows a comparison of the NN distribution between background-subtracted data and the expected signal for $m_H$=165~GeV hypothesis.  The background prediction and its uncertainties have been determined from the fit to data under the background-only hypothesis.  The LLR distribution as a function of $m_H$ is shown in Fig.~\ref{fig:results}(b) demonstrating the overall consistency of the data with the background-only hypothesis in the full $m_H$ range considered.
Table~\ref{tab:alllimitbay} and Fig.~\ref{fig:results}(c) present the expected and observed
upper limits as a ratio to the expected SM cross section.
Assuming $m_H$=165~GeV, the observed (expected) upper limit at 95\% C.L. on Higgs boson
production is a factor of 1.55 (1.36) times the SM cross section, representing
an improvement in sensitivity of over a factor of 6 relative to our previous 
publication~\cite{hohlfeld}, larger than expected from the luminosity increase alone.

Auxiliary material is provided in~\cite{EPAPS}.

\begin{acknowledgments}
%
We thank the staffs at Fermilab and collaborating institutions, 
and acknowledge support from the 
DOE and NSF (USA);
CEA and CNRS/IN2P3 (France);
FASI, Rosatom and RFBR (Russia);
CNPq, FAPERJ, FAPESP and FUNDUNESP (Brazil);
DAE and DST (India);
Colciencias (Colombia);
CONACyT (Mexico);
KRF and KOSEF (Korea);
CONICET and UBACyT (Argentina);
FOM (The Netherlands);
STFC and the Royal Society (United Kingdom);
MSMT and GACR (Czech Republic);
CRC Program, CFI, NSERC and WestGrid Project (Canada);
BMBF and DFG (Germany);
SFI (Ireland);
The Swedish Research Council (Sweden);
and
CAS and CNSF (China).

\end{acknowledgments}

\appendix

\begin{table*}
\begin{center}
{\Large{\bf Auxiliary material}}
\end{center}
\end{table*}

\setcounter{figure}{0}
\begin{table*}
\begin{flushleft}
{Figures \ref{fig:dphi_marginal} - \ref{fig:mtmin_marginal} are the distributions of variables used to define the final selection: $\Delta\phi(\ell,\ell)$, {\etmiss}, {\etmisssc}, and {\mtmin}.  The distributions are shown at final selection having removed final selection requirement on the plotted variable.  Figures \ref{fig:M} - \ref{fig:metscal} are the distributions of variables input into the neural network at final selection.}
\end{flushleft}
\end{table*}

\setcounter{table}{0}

\begin{table*}[hbt]
\begin{center}
\caption{\label{tab:yields_NN0.9} Expected and observed event yields in
each channel after final selection and requiring NN output $>$ 0.9.  The systematic uncertainty after fitting is shown.
}
\begin{tabular}{c|c|c|c}
\hline \hline
\multicolumn{4}{c}{NN output $>$ 0.9}\\\hline
               & $e^{\pm}\mu^{\mp}$&                $e^+e^-$      &         $\mu^+\mu^-$\\
               &final selection& final selection        & final selection \\
\hline
Signal ($m_H=165$~GeV) & 1.9 $\pm$ 0.2& 4.0 $\pm$ 0.4& 2.6 $\pm$ 0.3\\\hline
Total background & 3.2 $\pm$ 0.1 & 28.1 $\pm$ 0.5&11.6 $\pm$ 0.3\\\hline
\hline
Data &  3 & 30 & 11\\ 
\hline\hline
\end{tabular}
\end{center}
\end{table*}

\begin{table*}[hbt]
\begin{center}
\caption{\label{tab:higgs_cs} The production cross sections for the SM Higgs boson assumed for estimation of signal yield and the branching fraction for $H\rightarrow W^+W^-$.  The gluon fusion Higgs production cross section ($\sigma_{gg}\rightarrow H$) has been determined with a NNLO calculation while all other Higgs production cross sections ($\sigma(WH)$, $\sigma(ZH)$, and $\sigma(VBF)$) were determined with a NLO calculation.}
\begin{tabular*}{0.8\textwidth}{@{\extracolsep{\fill}}c|cccc|c}
\hline\hline
$m_H $ &$\sigma({gg}\rightarrow H)$ & $\sigma(WH)$  & $\sigma(ZH)$  &  $\sigma(VBF)$  & $B(H\rightarrow  W^+W^-)$\\
(GeV ) &  (fb) &  (fb)& (fb) & (fb)  & (\%)\\\hline
115	    &1240   &178.8  &107.4&	 79.1  &	       7.974		  \\            
120	    &1093   &152.9  &92.7	& 71.6  &	       13.20		  \\            
125	    &967    &132.4  &81.1	 &67.4  &	      20.18 \\            
130	    &858    &114.7  &70.9	 &62.5&  	      28.69		  \\            
135	    &764    &99.3&	  62.0&	 57.6  &	       38.28		  \\            
140	    &682    &86.0&	  54.2&	 52.6  &	       48.33		  \\            
145	    &611    &75.3&	  48.0&	 49.2  &	      58.33		 \\             
150	    &548    &66.0&	  42.5&	 45.7  &	      68.17		  \\            
155	    &492    &57.8&	  37.6&	 42.2  &	      78.23		  \\            
160	    &439    &50.7&	  33.3&	 38.6  &	      90.11		  \\            
165	    &389    &44.4&	  29.5&	 36.1  &	       96.10		  \\            
170	    &349    &38.9&	  26.1&	 33.6  &	      96.53		  \\            
175	    &314    &34.6&	  23.3&	 31.1  &	       95.94		  \\            
180	    &283    &30.7&	  20.8&	 28.6  &	       93.45		  \\            
185	    &255    &27.3&	  18.6&	 26.8  &	  83.79		  \\            
190	    &231    &24.3&	  16.6&	 24.9  &	      77.61		  \\            
195	    &210   & 21.7&	  15.0&	 23.0  &74.95		\\              
200	    &192&    19.3&	  13.5&   21.2  	& 73.47\\            
\hline\hline
\end{tabular*}
\end{center}
\end{table*}

\clearpage

\begin{table*}[hbt]
\begin{center}                                                                                                          
\caption{Systematic uncertainties in percent for the Monte Carlo samples and the multijet estimate.  Uncertainties are identical across all channels except where noted.  The nature of the uncertainty is indicated in the last column of the table.  Uncertainties that change the differential distribution of the final discriminant are labeled with ``D", while uncertainties that affect only the normalization are indicated by ``N".  The values for uncertainties with a differential dependence
correspond to the maximum amplitude of fluctuations in the final discriminant.}
\begin{ruledtabular}
\begin{tabular}{ l  c  c  c  c  c  c c c}
                 & $\Sigma$ Bkgd  & Signal & $Z+jets/\gamma$ & $W+jets/\gamma$ & $t\bar{t}$  & Diboson &   Multijet & Nature\\
\hline
Lepton identification& $\pm$4&$\pm$4&$\pm$4&$\pm$4&$\pm$4&$\pm$4&-& N\\
Lepton momentum resolution& $\pm$2&$\pm$2&$\pm$1&$\pm$1&$\pm$1&$\pm$2&-&D\\
Jet energy scale & $\pm$4 & $\pm$1 &$\pm$8 &$\pm$1 &$\pm$1 &$\pm$1 &-& D \\
Jet energy resolution &$\pm$3 &$\pm$1 &$\pm$4 &$\pm$2 &$\pm$1 &$\pm$1 &-&D\\
Jet identification   &$\pm$4 &$\pm$1 &$\pm$6 &$\pm$4 &$\pm $1 &$\pm$1 &-&D\\
$Z-p_T$ correction &$\pm$1&-&$\pm$3&-&-&-&-&D\\
$W-p_T$ correction &$\pm$1&-&-&$\pm$2&-&-&-&D\\
Diboson NLO correction & $\pm$1&$\pm$1&-&-&-&$\pm$1&-&D\\
Multijet Normalization  $e^+e^-$&  $\pm$2 &   &	- &	 -       &   -     & -   &   $\pm$20  & N\\
Multijet Normalization  $e^{\pm}\mu^{\mp}$&  $\pm$1 &   &	- &	 -       &   -     & -   &   $\pm$10  & N\\
Multijet Normalization  $\mu^+\mu^-$&  $\pm$2 &   &	- &	 -       &   -     & -   &   $\pm$20  & N\\
Cross section	   &      $\pm$7        &  $\pm$10	   &	$\pm$6	     &	 $\pm$6	       &   $\pm$10	     &	$\pm$6      &   -   & N\\
PDF		   &$\pm$ 1          &  $\pm$1   &	-	     &	 -	       &   -	     &	-      &   -   &N\\
Luminosity	   & $\pm$6.1          &  $\pm$6.1   &	-	     &	 -	       &   -	     &	-      &   -   &N\\
\end{tabular}
\end{ruledtabular}
\label{tab:systematics}
\end{center}
\end{table*}

\begin{figure*}[htp]
  \begin{center}
  \begin{tabular}{lr}
  \includegraphics[width=1.0\columnwidth]{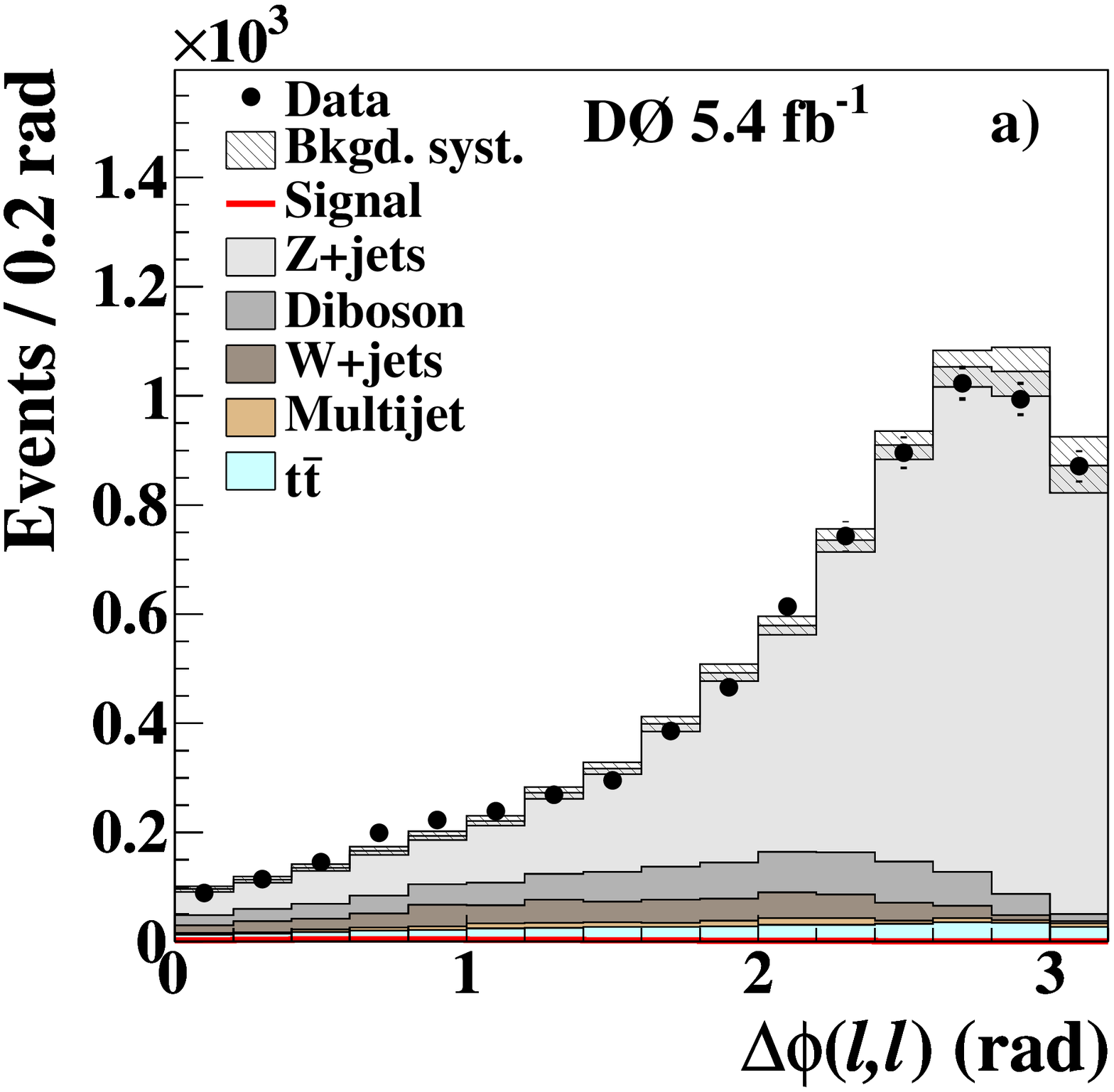} &
  \includegraphics[width=1.0\columnwidth]{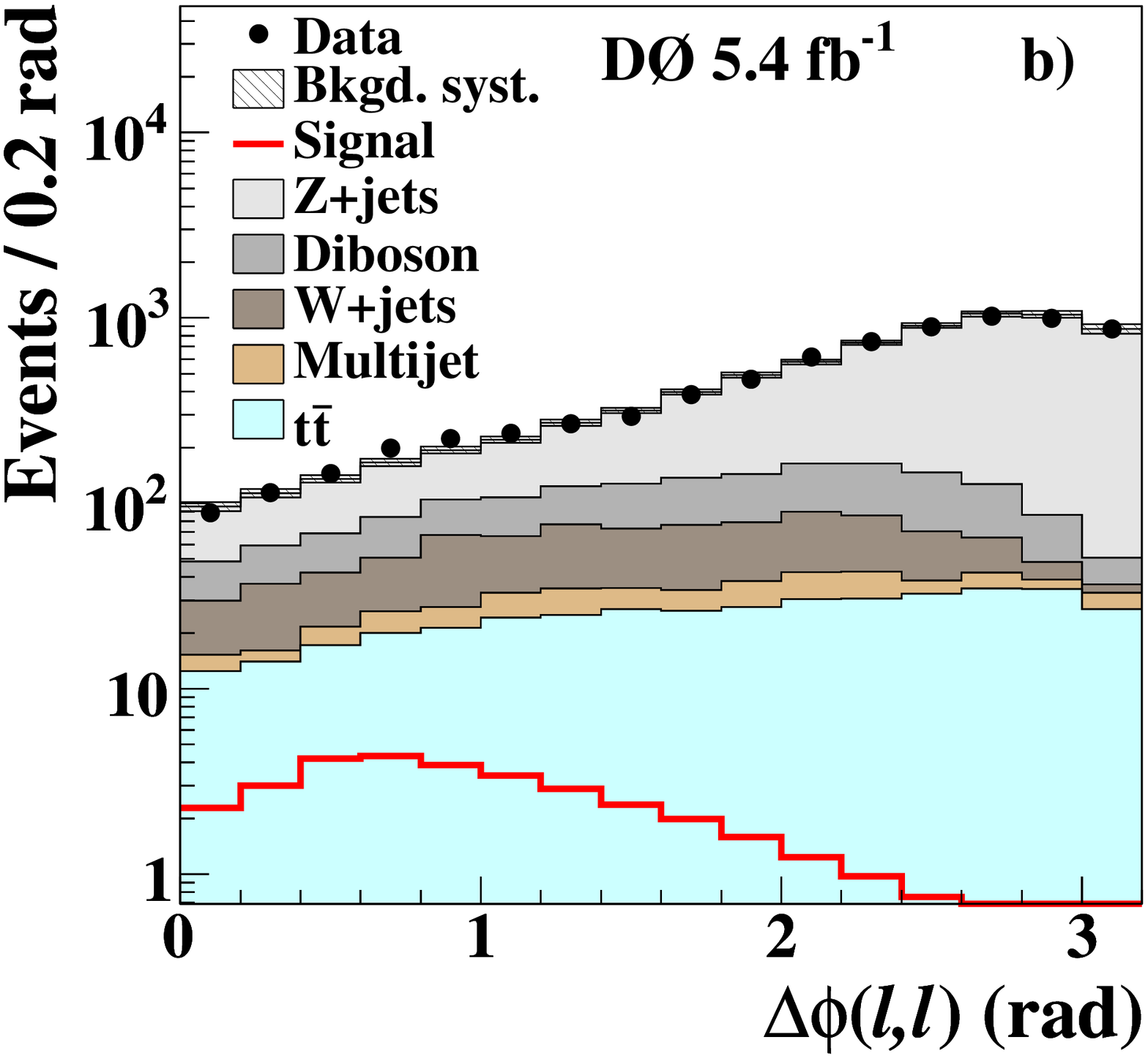} 
  \end{tabular}
  \end{center}
  \caption{The $\Delta\phi(\ell,\ell)$ angle between the two leptons after all final selection requirements except for the selection on $\Delta\phi(\ell,\ell)$ in linear (a) and logarithmic (b) scale for the combination of $e^+e^-$, $\mu^+\mu^-$, and $e^{\pm}\mu^{\mp}$ channels.  The signal is shown for $m_H$=165~GeV and is scaled to the SM prediction for the combination of Higgs boson production from gluon fusion, vector boson fusion, and associated production.  The systematic uncertainty is shown after fitting.
  \label{fig:dphi_marginal}}
\end{figure*}

\begin{figure*}[htp]
  \begin{center}
  \begin{tabular}{lr}
  \includegraphics[width=1.0\columnwidth]{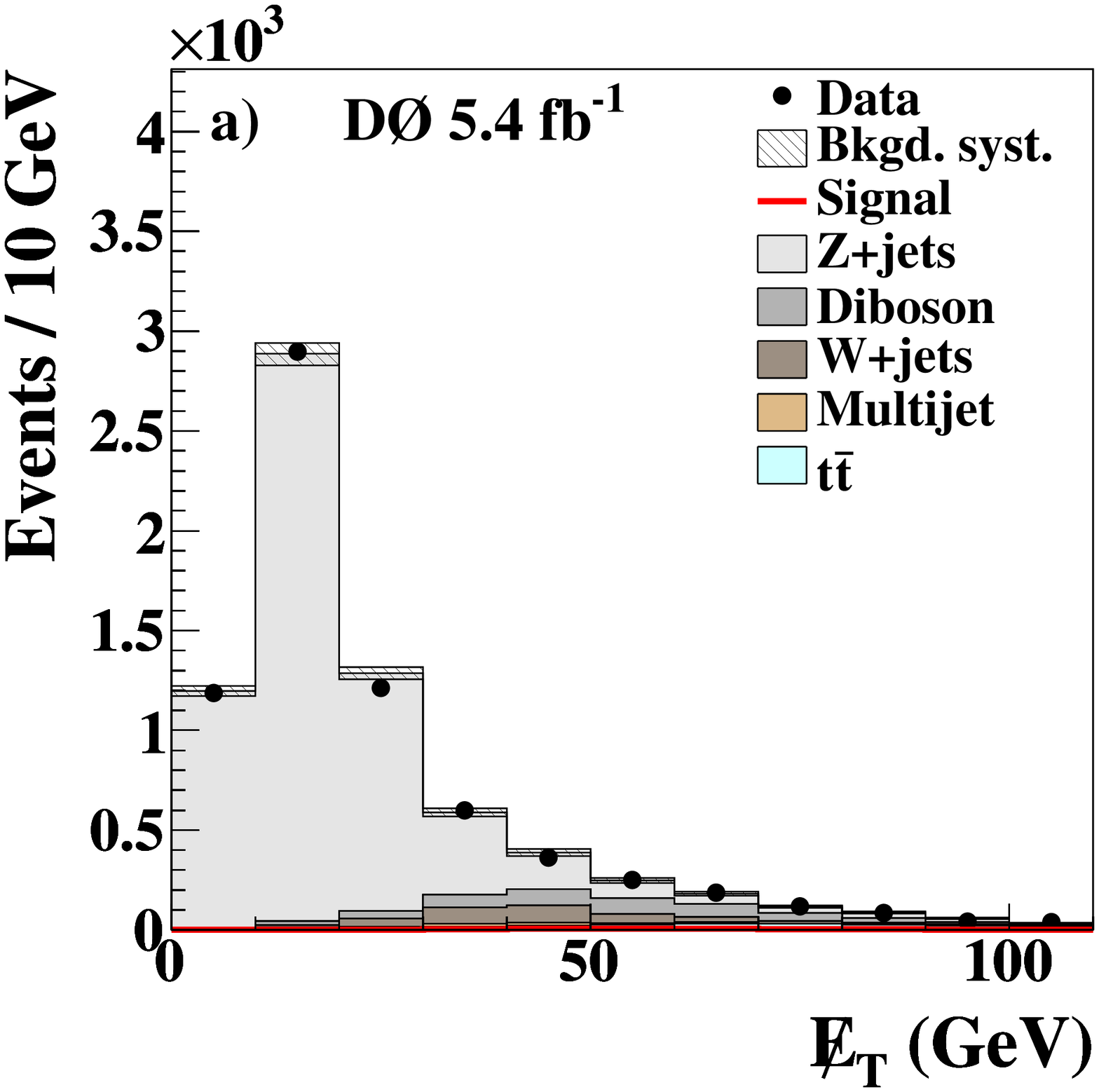} &
  \includegraphics[width=1.0\columnwidth]{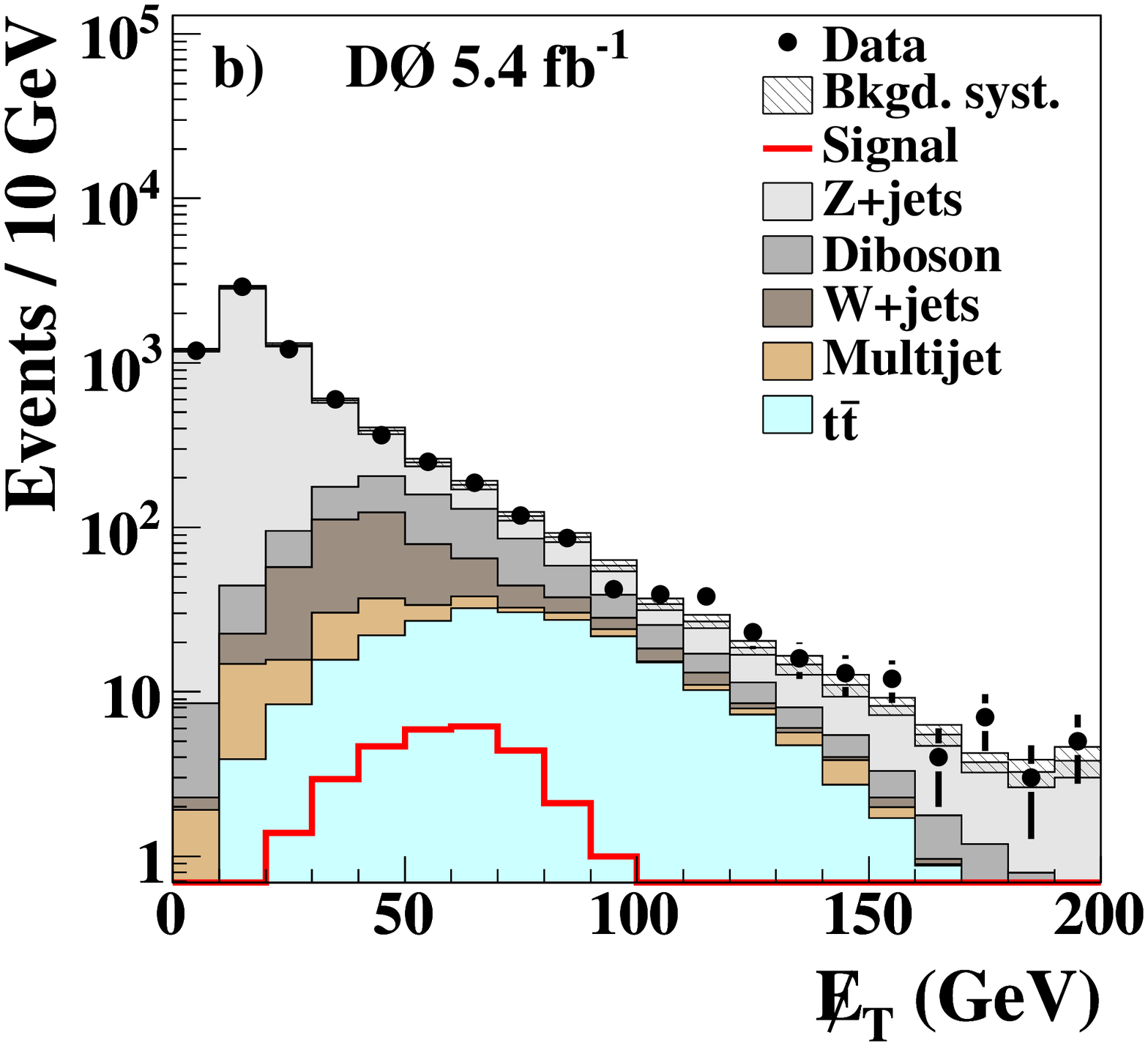} 
  \end{tabular}
  \end{center}
  \caption{The {\etmiss} after all selections except for the selection on {\etmiss} in linear (a) and logarithmic (b) scale for the combination of $e^+e^-$, $\mu^+\mu^-$, and $e^{\pm}\mu^{\mp}$ channels.  The signal is shown for $m_H$=165~GeV and is scaled to the SM prediction for the combination of Higgs boson production from gluon fusion, vector boson fusion, and associated production.  The systematic uncertainty is shown after fitting.
  \label{fig:met_marginal}}
\end{figure*}

\begin{figure*}[htp]
  \begin{center}
  \begin{tabular}{lcr}
  \includegraphics[width=1.0\columnwidth]{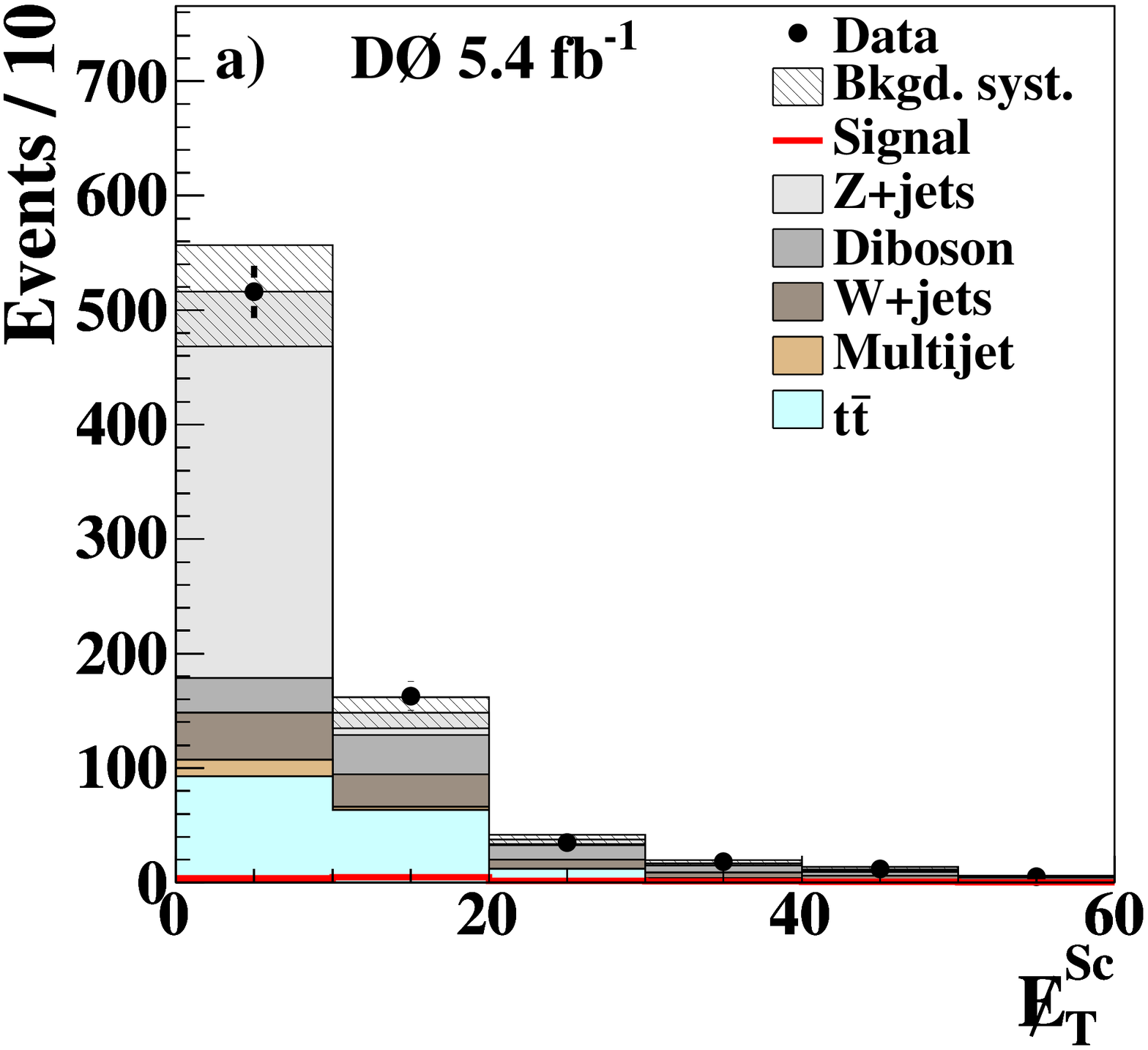} &
  \includegraphics[width=1.0\columnwidth]{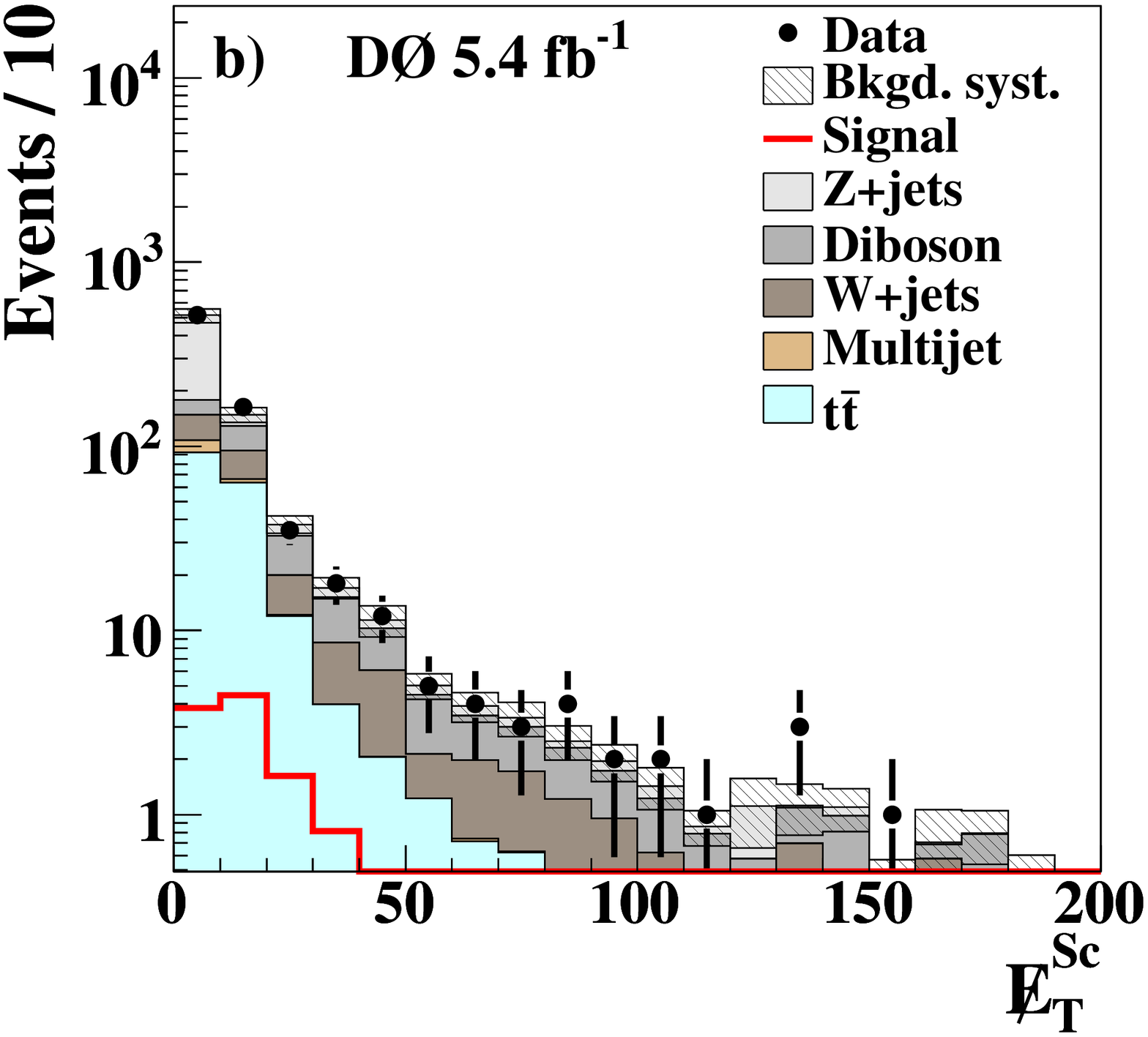} &
  \end{tabular}
  \end{center}
  \caption{The {\etmisssc} after all selections except for the selection on {\etmisssc} in linear (a) and logarithmic (b) scale for the combination of $e^+e^-$, and $e^{\pm}\mu^{\mp}$ channels.  The signal is shown for $m_H$=165~GeV and is scaled to the SM prediction for the combination of Higgs boson production from gluon fusion, vector boson fusion, and associated production.  The systematic uncertainty is shown after fitting.
  \label{fig:metscale_marginal}}
\end{figure*}

\begin{figure*}[htp]
  \begin{center}
  \begin{tabular}{lcr}
  \includegraphics[width=1.0\columnwidth]{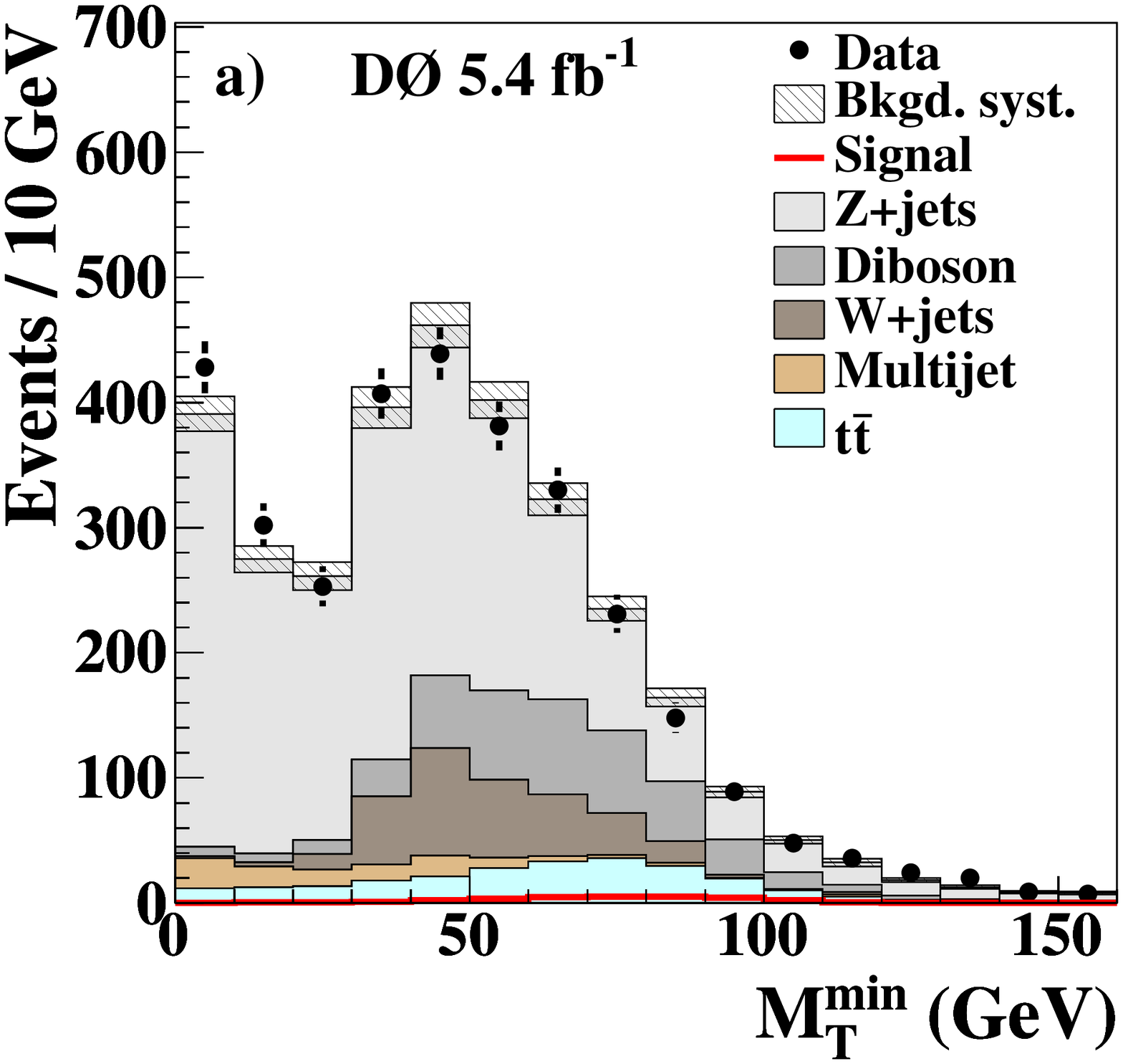} &
  \includegraphics[width=1.0\columnwidth]{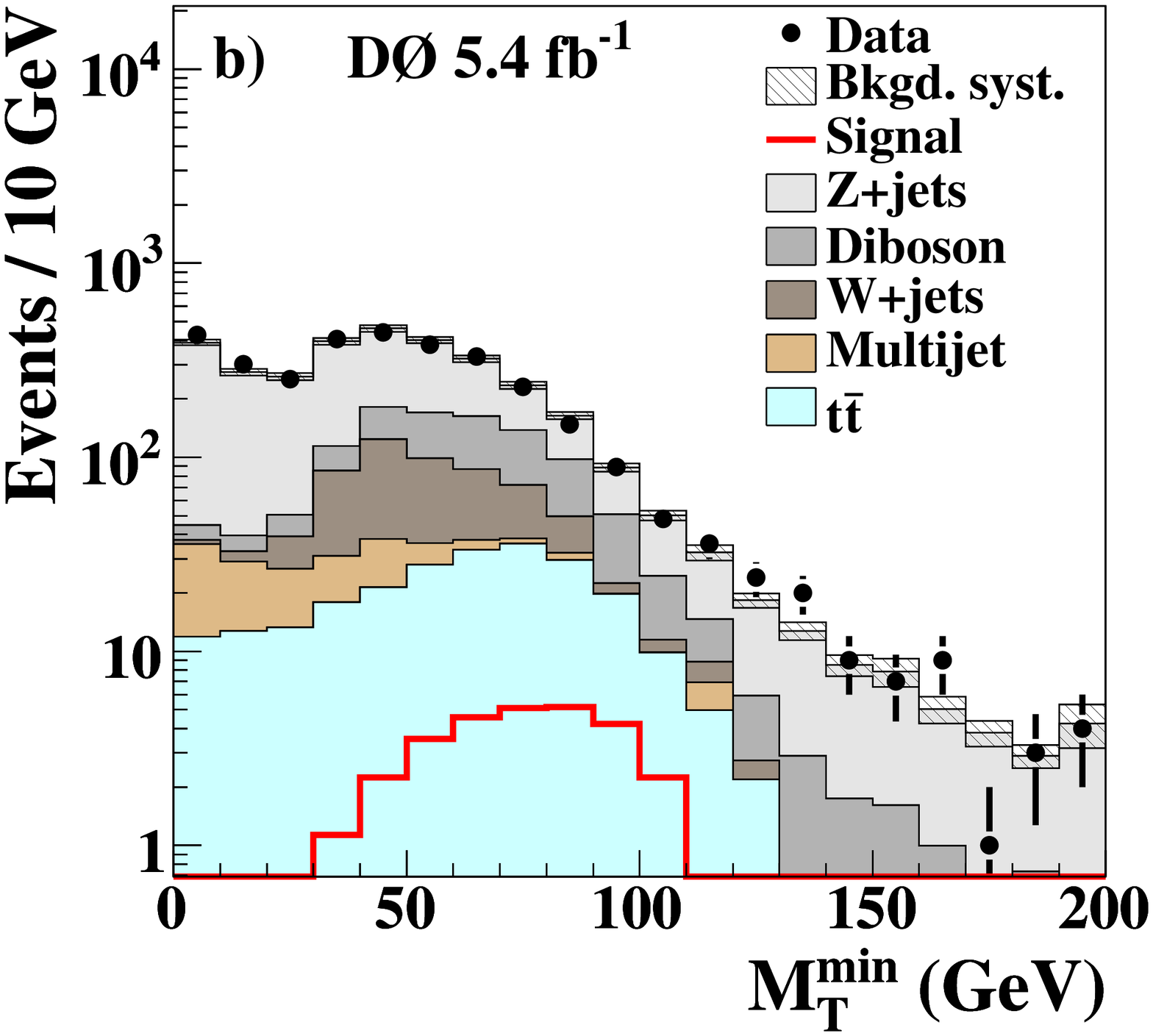} &
  \end{tabular}
  \end{center}
  \caption{The \mtmin after all selections except for the selection on \mtmin ~in linear (a) and logarithmic (b) scale for the combination of $e^+e^-$, $\mu^+\mu^-$, and $e^{\pm}\mu^{\mp}$ channels.  The signal is shown for $m_H$=165~GeV and is scaled to the SM prediction for the combination of Higgs boson production from gluon fusion, vector boson fusion, and associated production.  The systematic uncertainty is shown after fitting.
  \label{fig:mtmin_marginal}}
\end{figure*}

\begin{figure*}[htp]
  \begin{center}
  \begin{tabular}{lcr}
  \includegraphics[width=1.0\columnwidth]{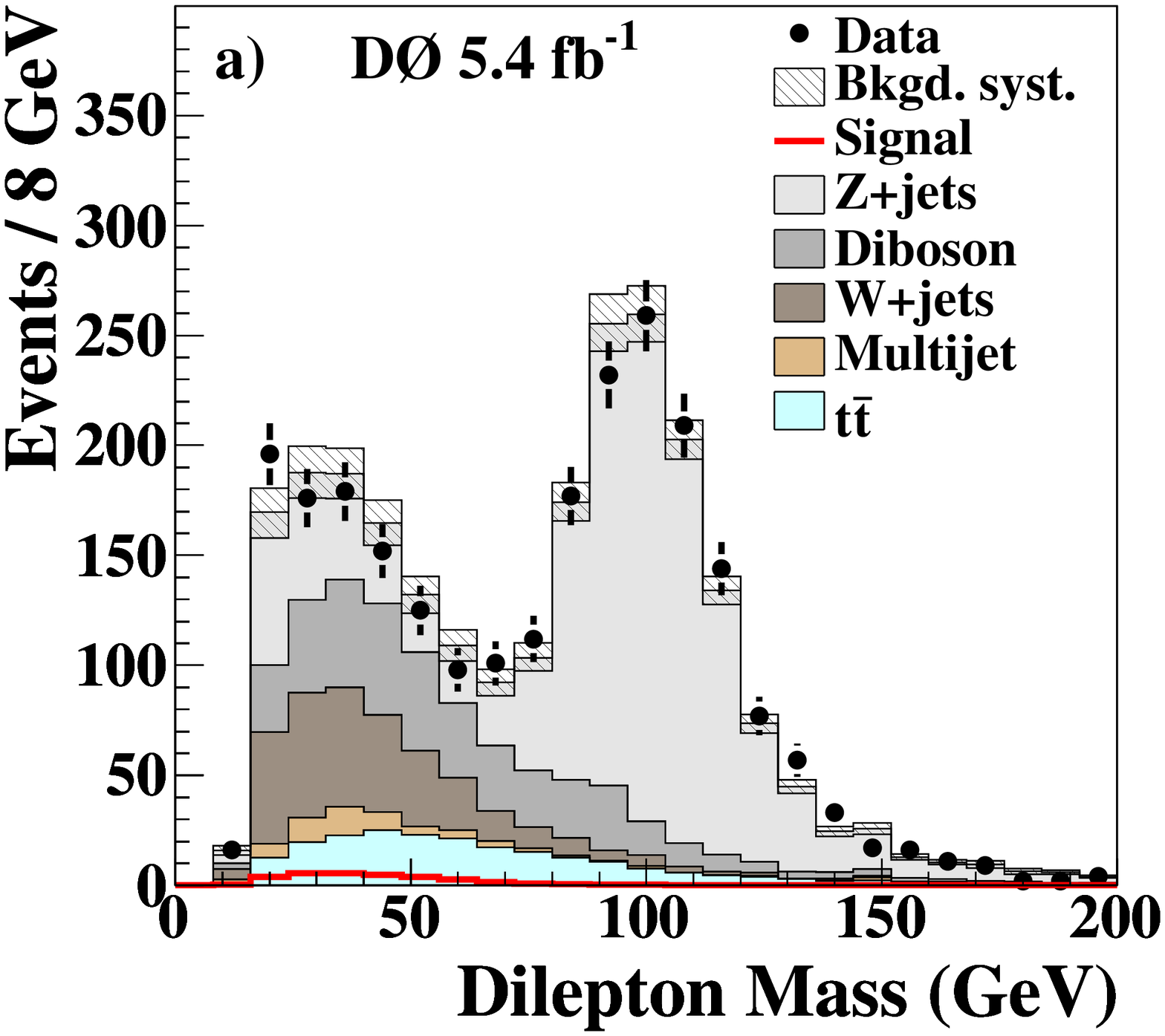} &
  \includegraphics[width=1.0\columnwidth]{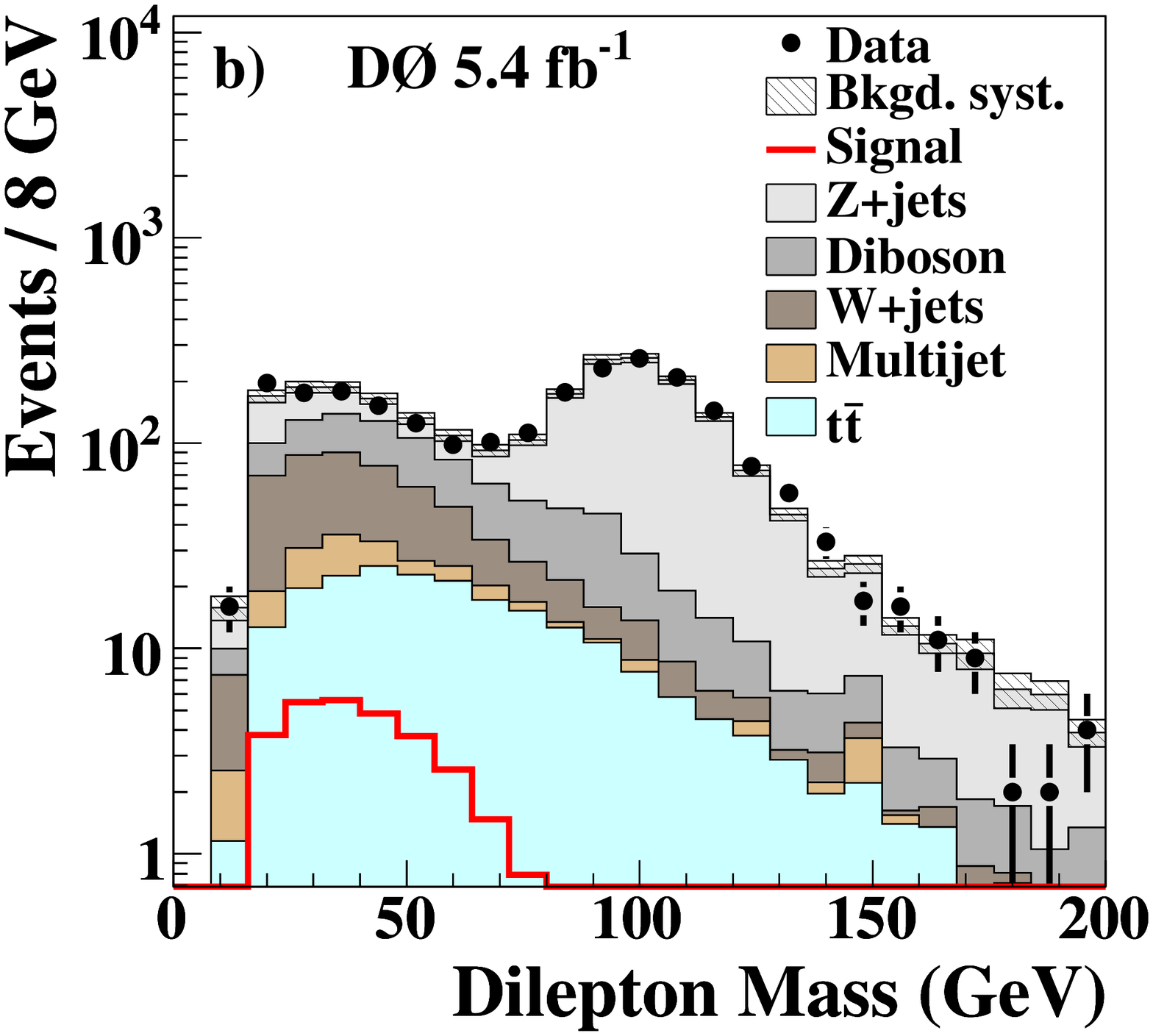} &
  \end{tabular}
  \end{center}
  \caption{The dilepton invariant mass at final selection in linear (a) and logarithmic (b) scale for the combination of $e^+e^-$, $\mu^+\mu^-$, and $e^{\pm}\mu^{\mp}$ channels.  The signal is shown for $m_H$=165~GeV and is scaled to the SM prediction for the combination of Higgs boson production from gluon fusion, vector boson fusion, and associated production.  The systematic uncertainty is shown after fitting.
  \label{fig:M}}
\end{figure*}

\begin{figure*}[htp]
  \begin{center}
  \begin{tabular}{lcr}
  \includegraphics[width=1.0\columnwidth]{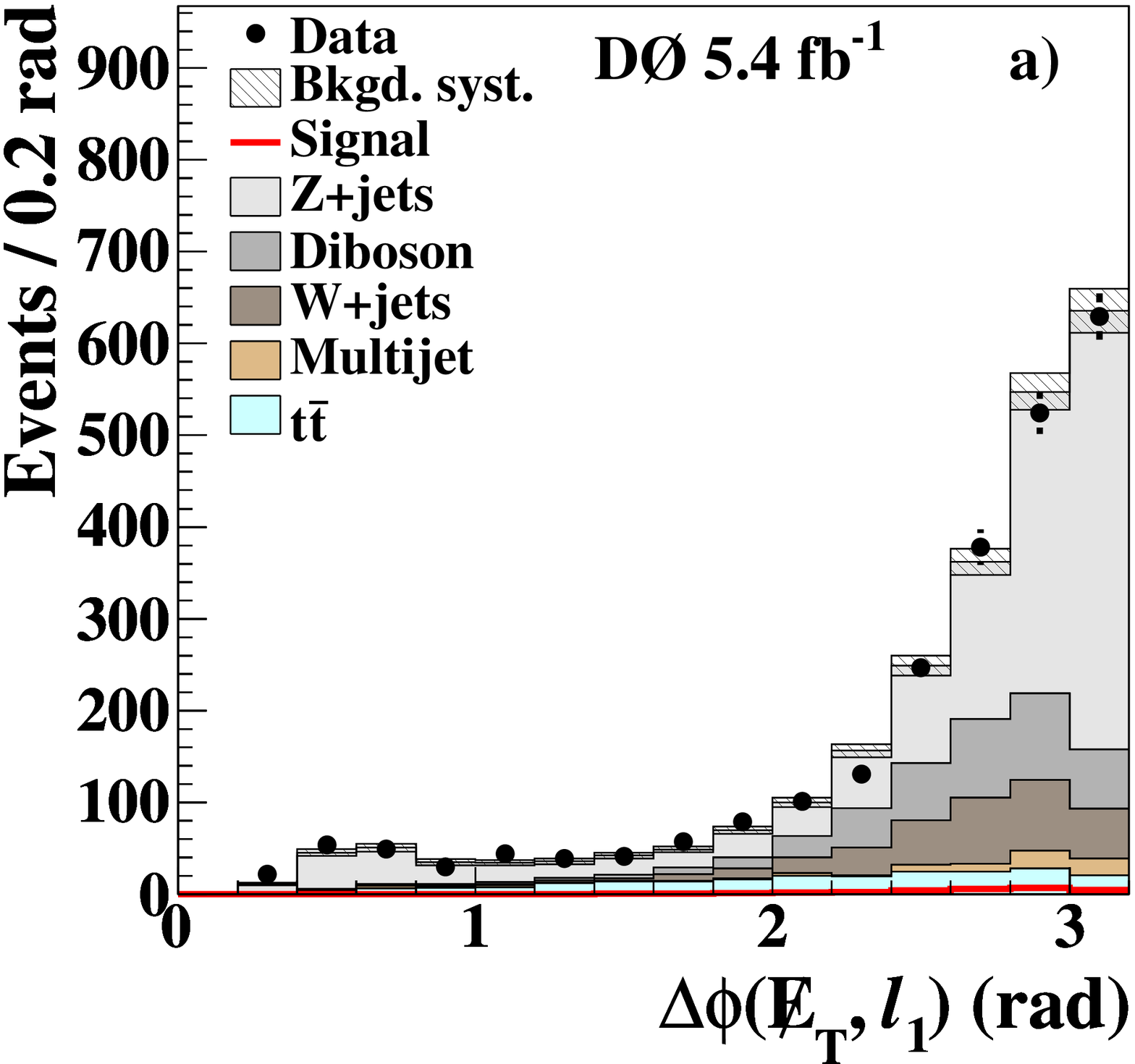} &
  \includegraphics[width=1.0\columnwidth]{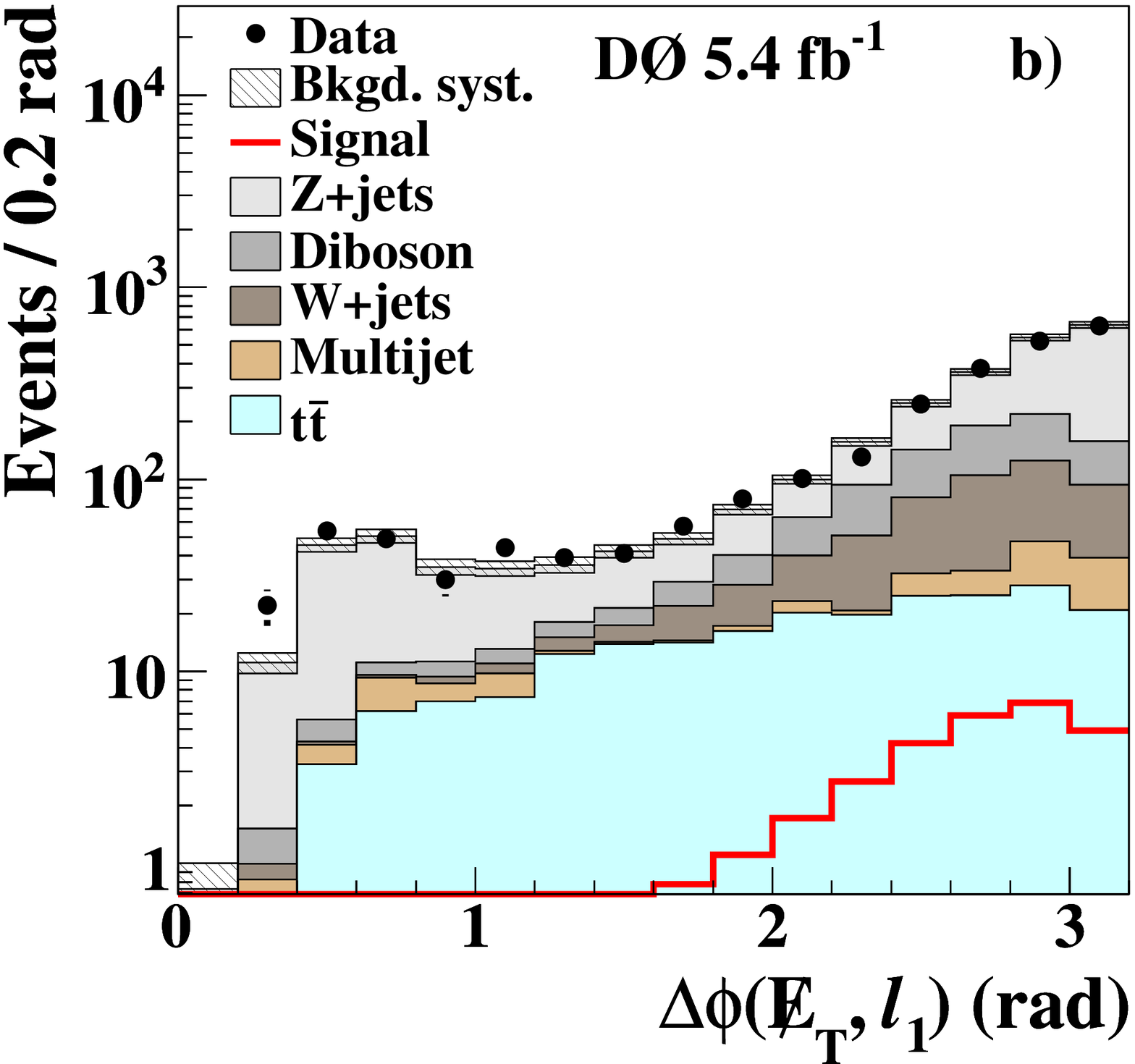} &
  \end{tabular}
  \end{center}
  \caption{The $\Delta\phi(\etmiss,\ell_1)$ at final selection in linear (a) and logarithmic (b) scale for the combination of $e^+e^-$, $\mu^+\mu^-$, and $e^{\pm}\mu^{\mp}$ channels.  The signal is shown for $m_H$=165~GeV and is scaled to the SM prediction for the combination of Higgs boson production from gluon fusion, vector boson fusion, and associated production.  The systematic uncertainty is shown after fitting.
  \label{fig:dphimetl1}}
\end{figure*}

\begin{figure*}[htp]
  \begin{center}
  \begin{tabular}{lcr}
  \includegraphics[width=1.0\columnwidth]{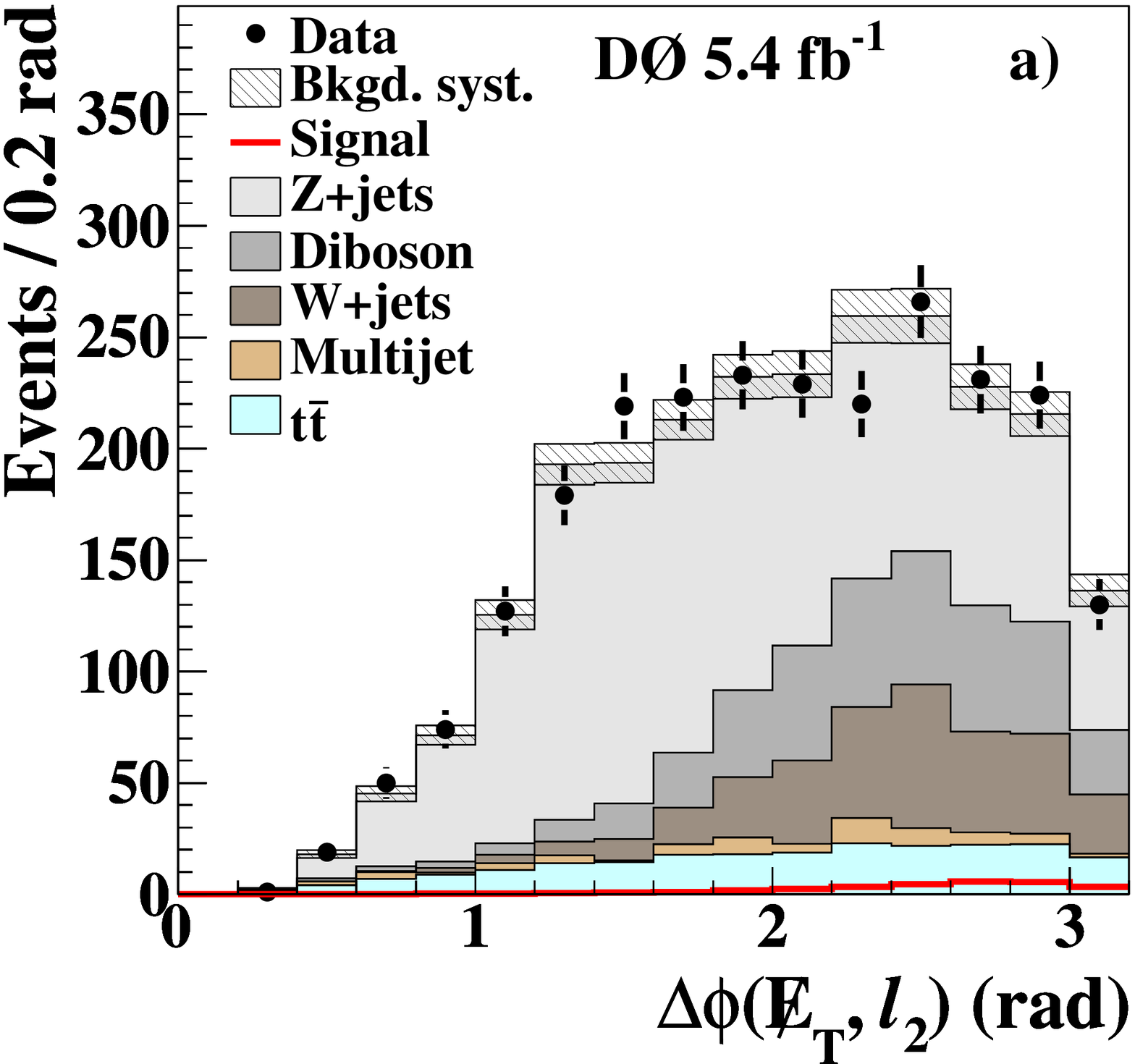} &
  \includegraphics[width=1.0\columnwidth]{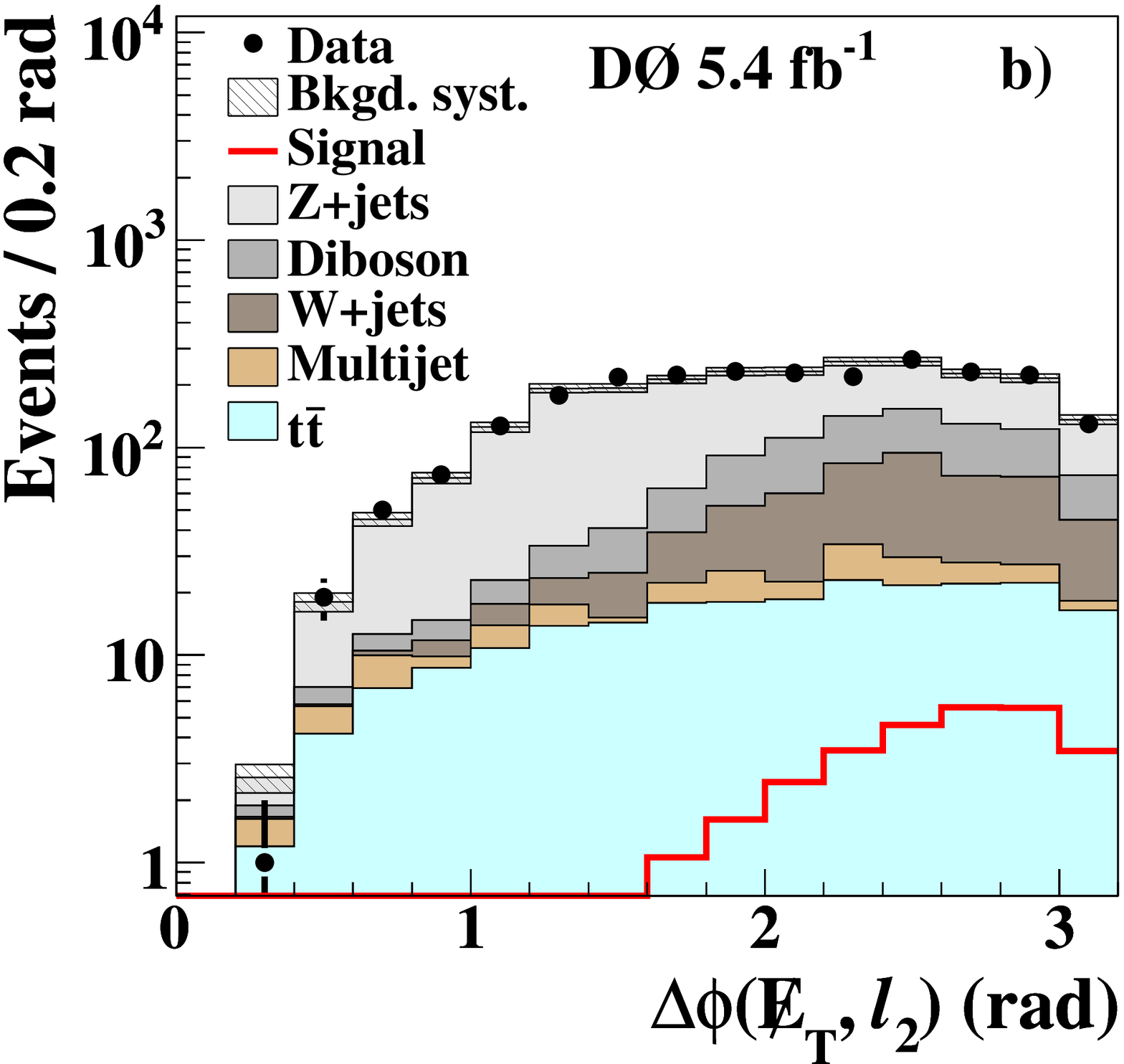} &
  \end{tabular}
  \end{center}
  \caption{The $\Delta\phi(\etmiss,\ell_2)$ at final selection in linear (a) and logarithmic (b) scale for the combination of $e^+e^-$, $\mu^+\mu^-$, and $e^{\pm}\mu^{\mp}$ channels.  The signal is shown for $m_H$=165~GeV and is scaled to the SM prediction for the combination of Higgs boson production from gluon fusion, vector boson fusion, and associated production.  The systematic uncertainty is shown after fitting.
  \label{fig:dphimetl2}}
\end{figure*}

\begin{figure*}[htp]
  \begin{center}
  \begin{tabular}{lcr}
  \includegraphics[width=1.0\columnwidth]{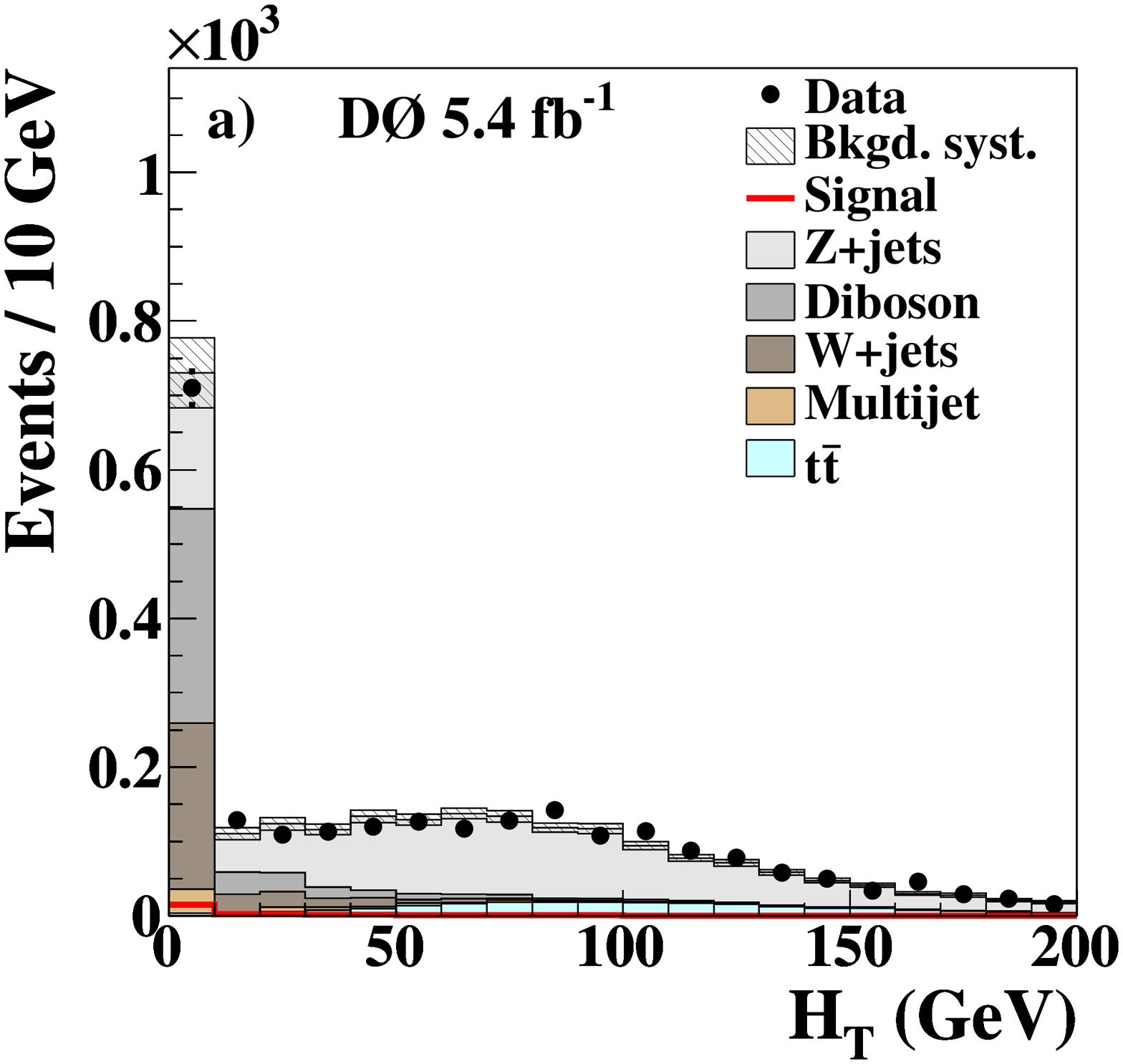} &
  \includegraphics[width=1.0\columnwidth]{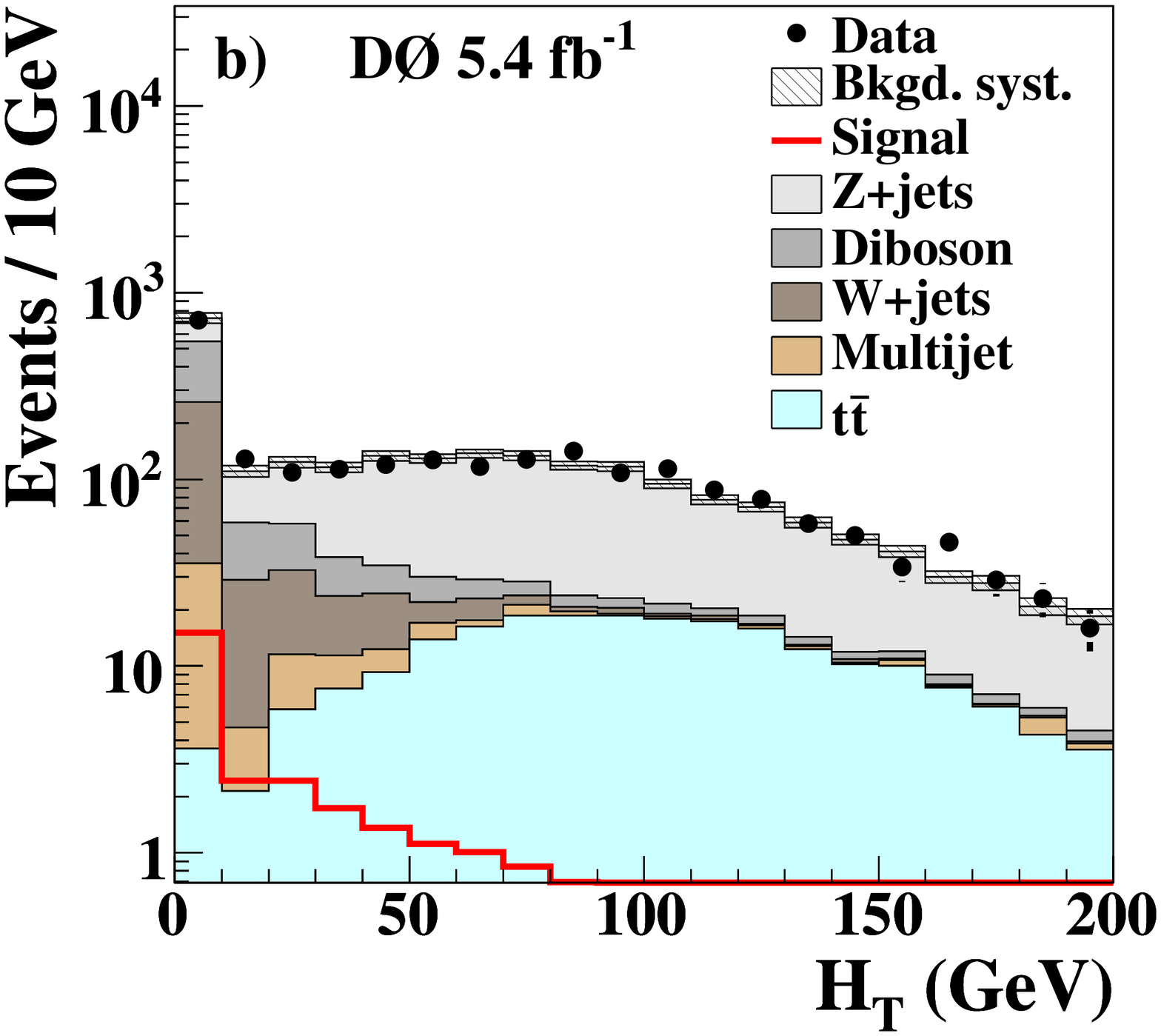} &
  \end{tabular}
  \end{center}
  \caption{The scalar sum of the transverse momenta of the jets at final selection in linear (a) and logarithmic (b) scale for the combination of $e^+e^-$, $\mu^+\mu^-$, and $e^{\pm}\mu^{\mp}$ channels.  The signal is shown for $m_H$=165~GeV and is scaled to the SM prediction for the combination of Higgs boson production from gluon fusion, vector boson fusion, and associated production.  The systematic uncertainty is shown after fitting.
  \label{fig:ht}}
\end{figure*}

\begin{figure*}[htp]
  \begin{center}
  \begin{tabular}{lcr}
  \includegraphics[width=1.0\columnwidth]{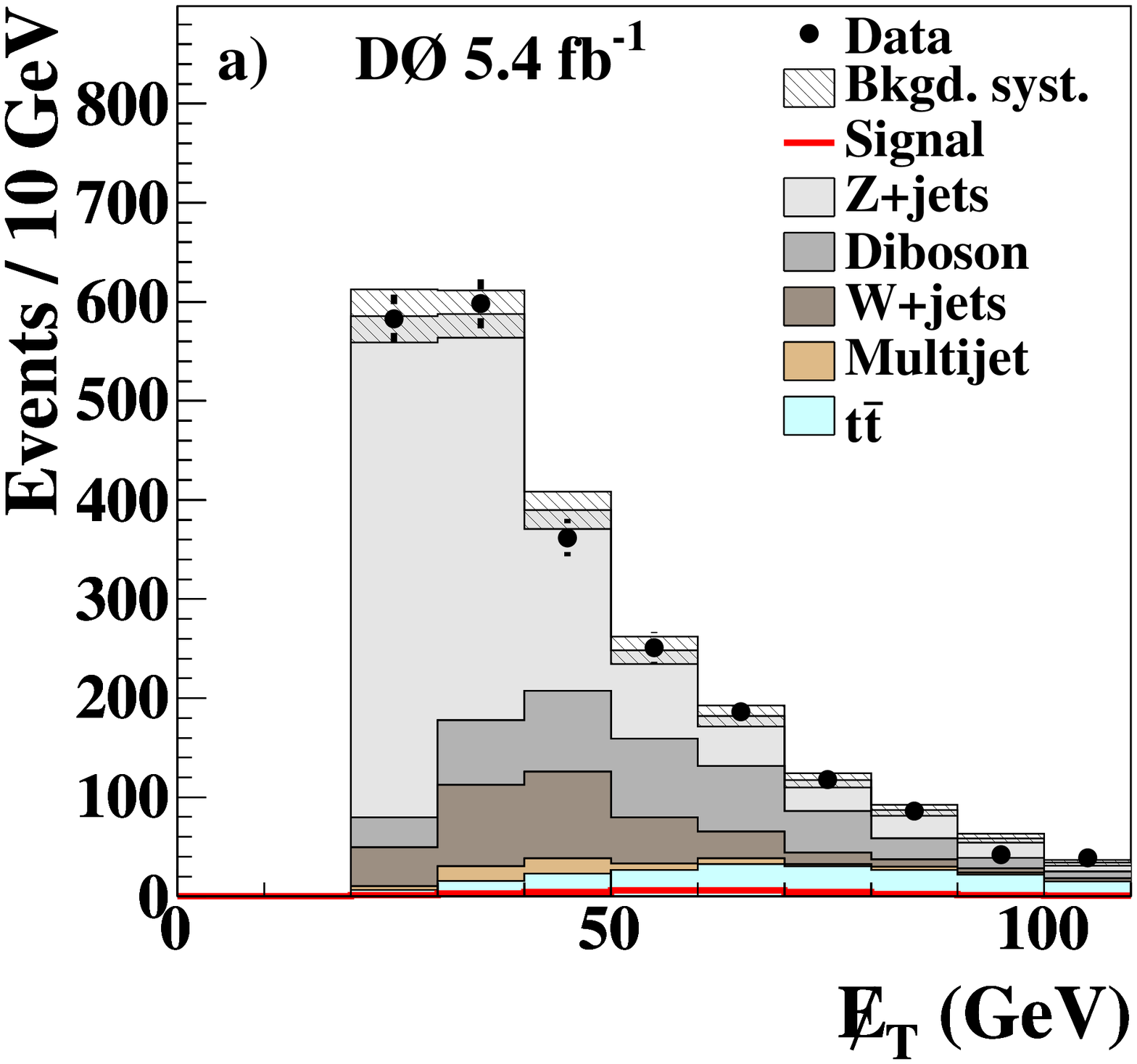} &
  \includegraphics[width=1.0\columnwidth]{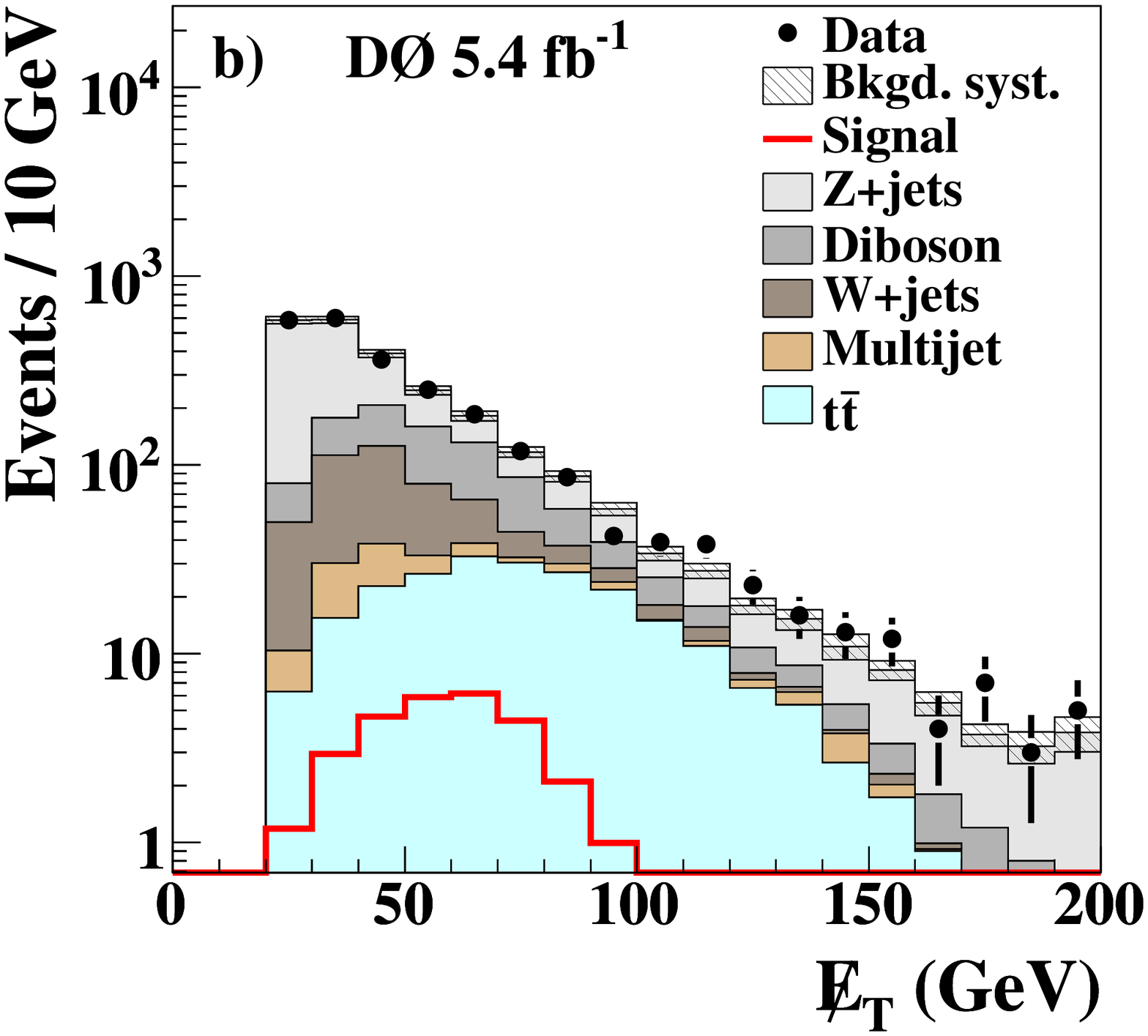} &
  \end{tabular}
  \end{center}
  \caption{The {\etmiss} at final selection in linear (a) and logarithmic (b) scale for the combination of $e^+e^-$, $\mu^+\mu^-$, and $e^{\pm}\mu^{\mp}$ channels.  The signal is shown for $m_H$=165~GeV and is scaled to the SM prediction for the combination of Higgs boson production from gluon fusion, vector boson fusion, and associated production.  The systematic uncertainty is shown after fitting.
  \label{fig:met}}
\end{figure*}

\begin{figure*}[htp]
  \begin{center}
  \begin{tabular}{lcr}
  \includegraphics[width=1.0\columnwidth]{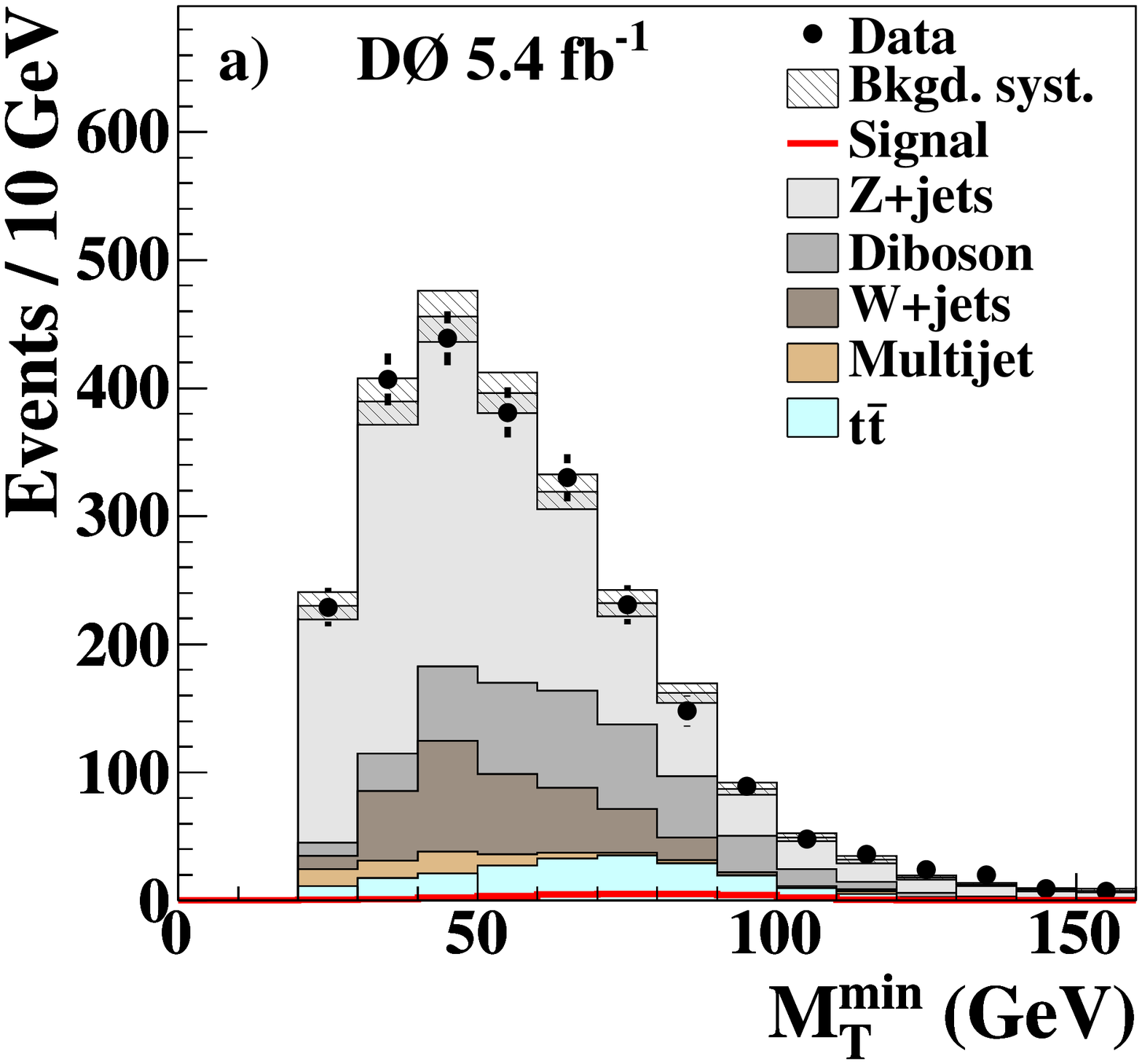} &
  \includegraphics[width=1.0\columnwidth]{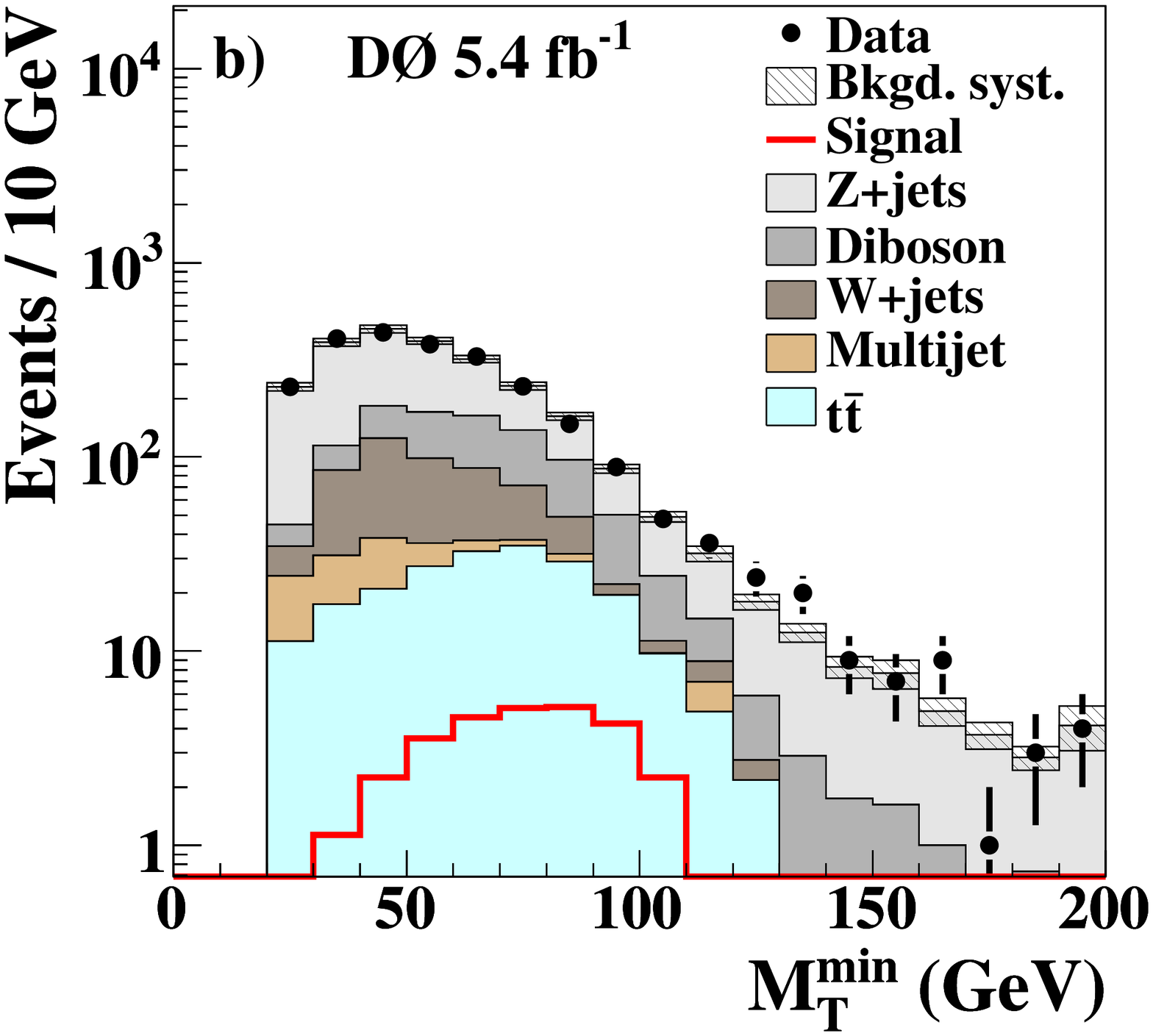} &
  \end{tabular}
  \end{center}
  \caption{The {\mtmin} ~at final selection in linear (a) and logarithmic (b) scale for the combination of $e^+e^-$, $\mu^+\mu^-$, and $e^{\pm}\mu^{\mp}$ channels.  The signal is shown for $m_H$=165~GeV and is scaled to the SM prediction for the combination of Higgs boson production from gluon fusion, vector boson fusion, and associated production.  The systematic uncertainty is shown after fitting.
  \label{fig:minmt}}
\end{figure*}

\begin{figure*}[htp]
  \begin{center}
  \begin{tabular}{lcr}
  \includegraphics[width=1.0\columnwidth]{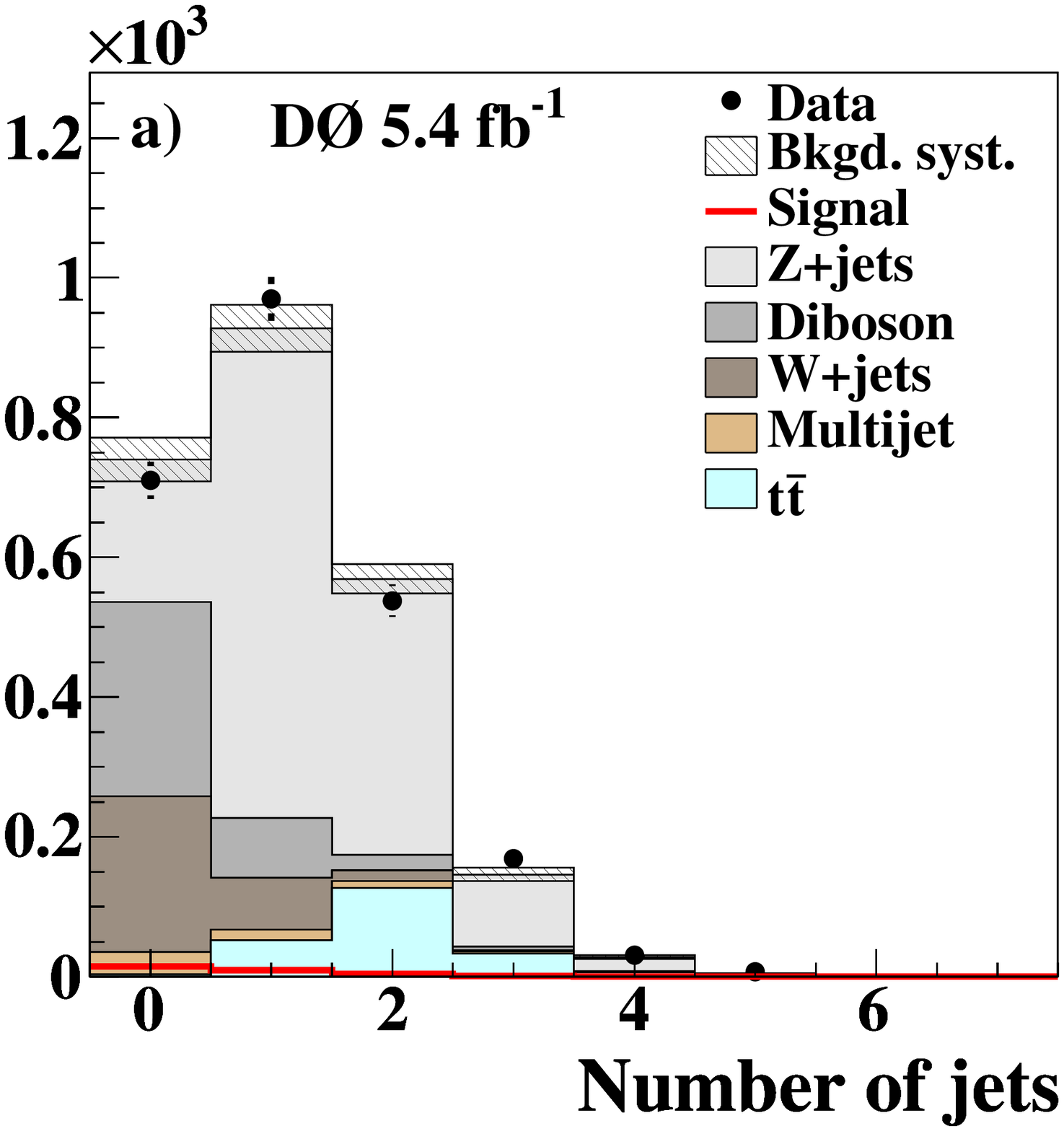} &
  \includegraphics[width=1.0\columnwidth]{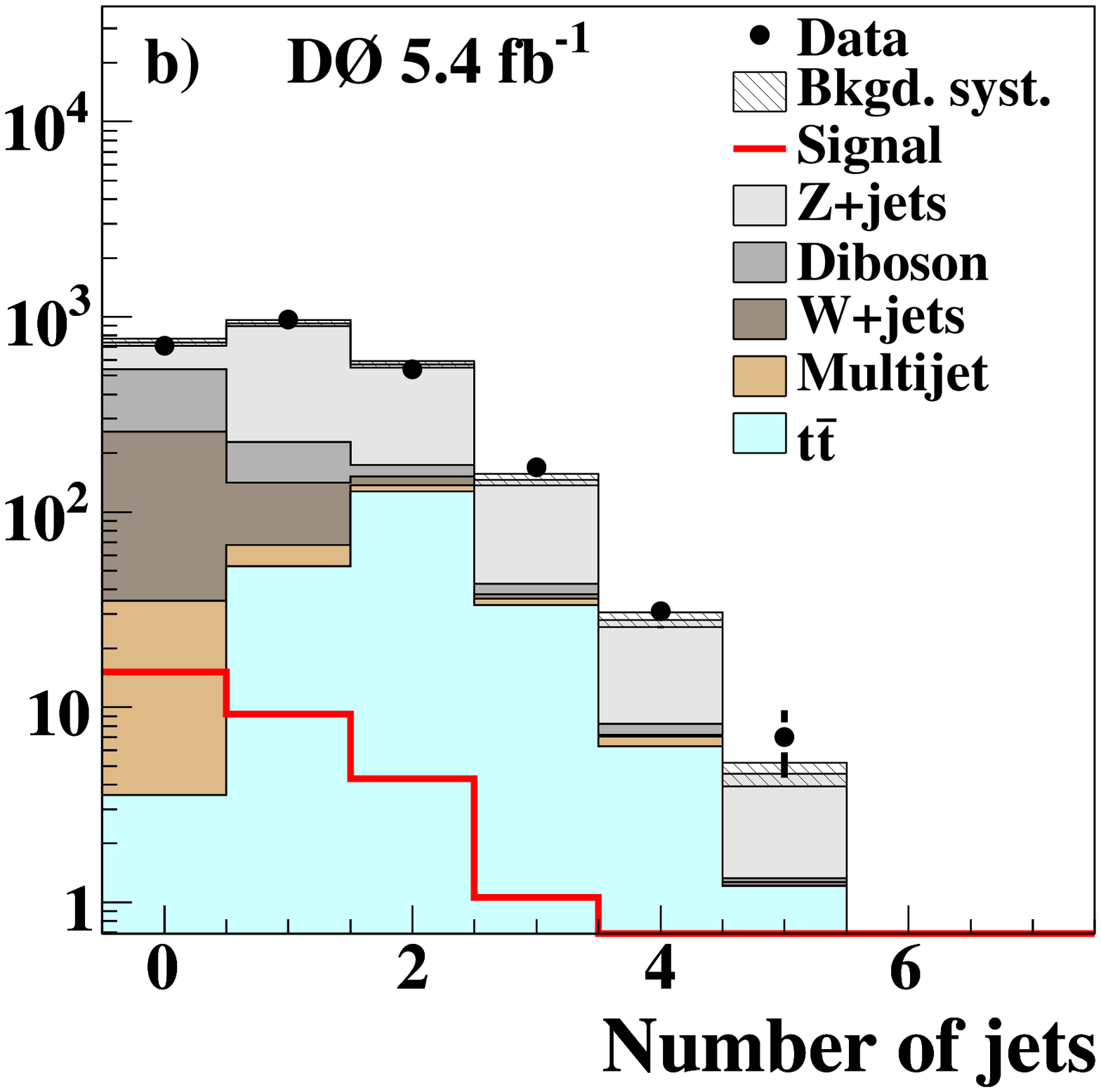} &
  \end{tabular}
  \end{center}
  \caption{The number of identified jets at final selection in linear (a) and logarithmic (b) scale for the combination of $e^+e^-$, $\mu^+\mu^-$, and $e^{\pm}\mu^{\mp}$ channels.  The signal is shown for $m_H$=165~GeV and is scaled to the SM prediction for the combination of Higgs boson production from gluon fusion, vector boson fusion, and associated production.  The systematic uncertainty is shown after fitting.
  \label{fig:njet}}
\end{figure*}

\begin{figure*}[htp]
  \begin{center}
  \begin{tabular}{lcr}
  \includegraphics[width=1.0\columnwidth]{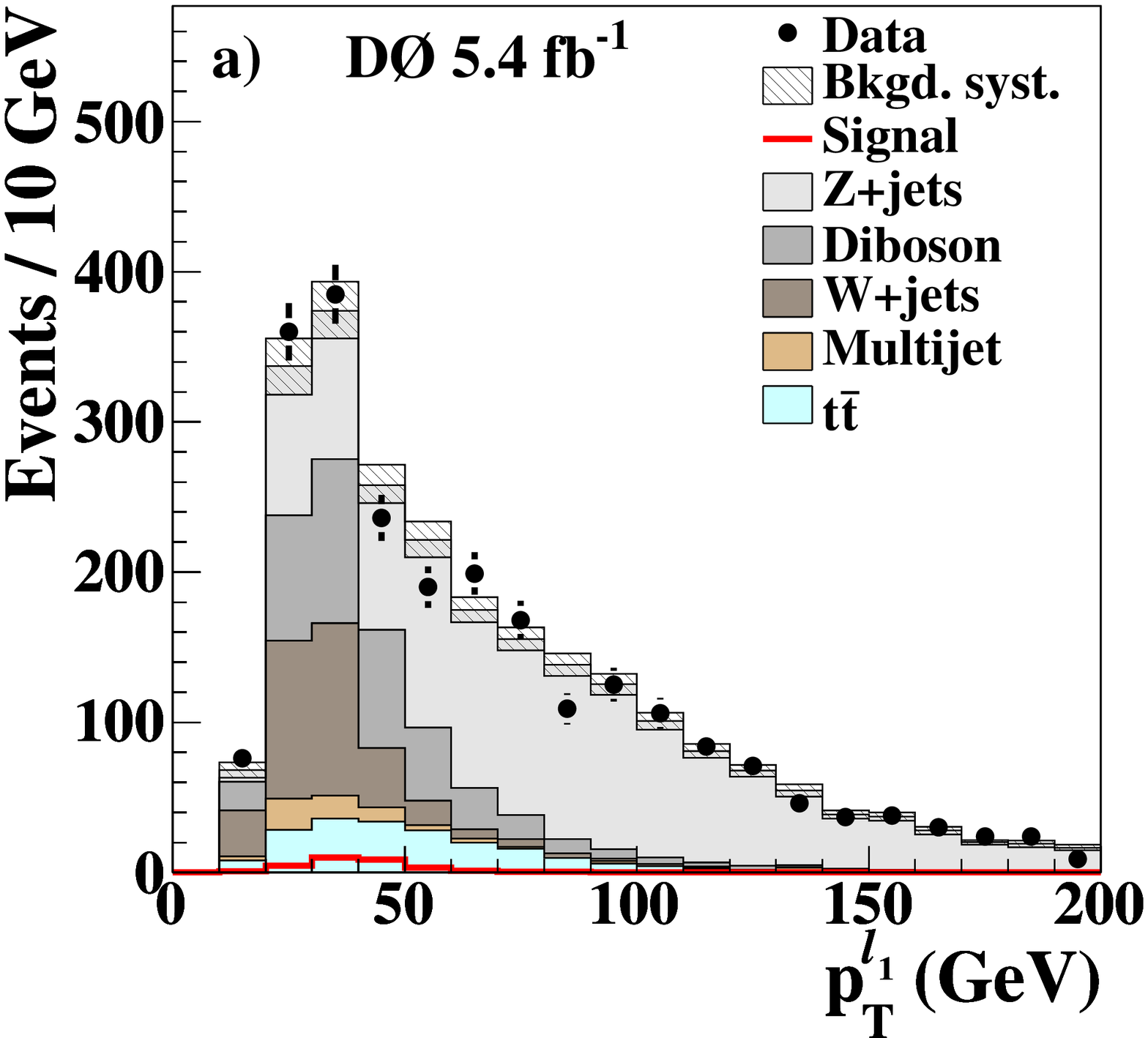} &
  \includegraphics[width=1.0\columnwidth]{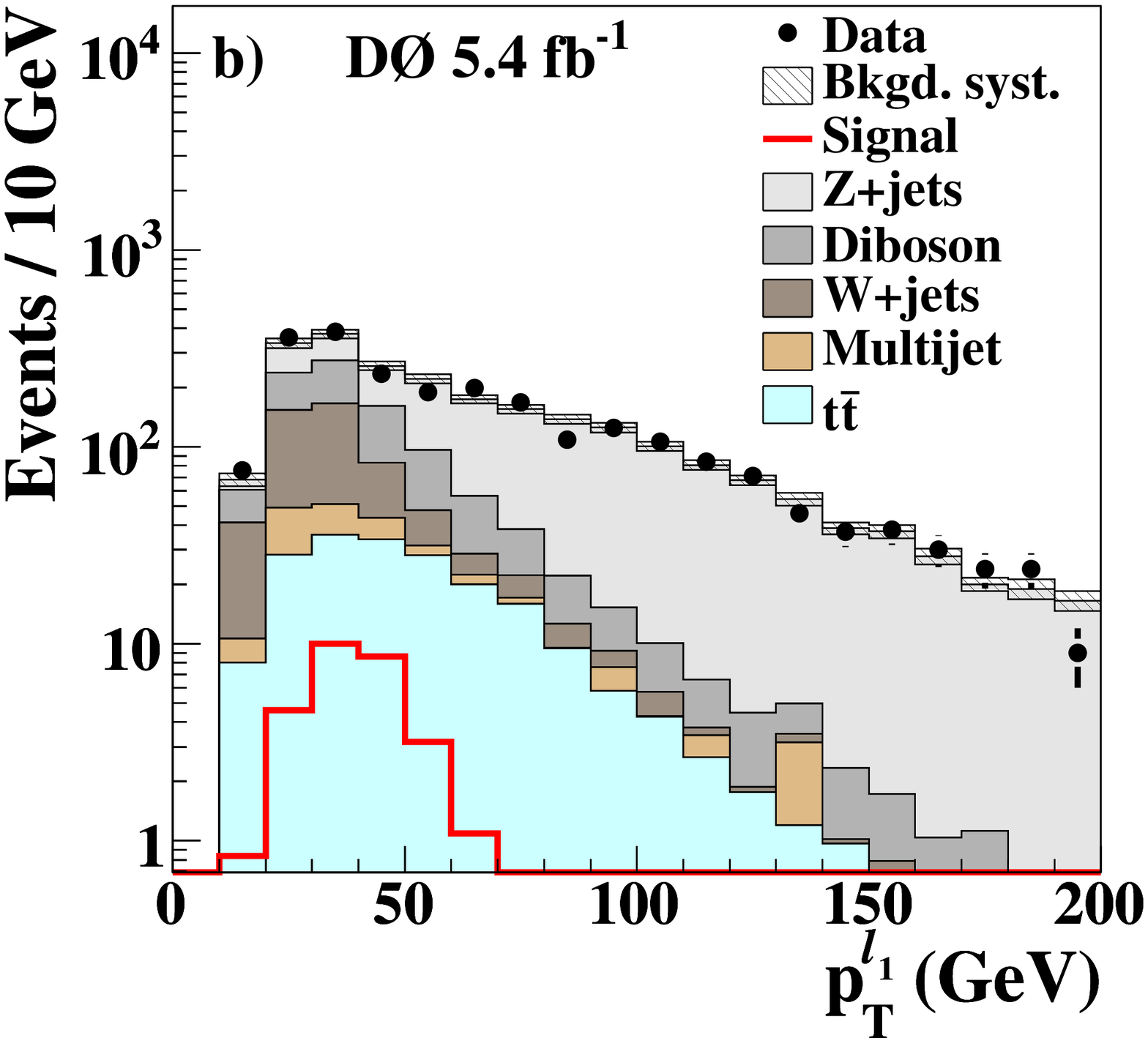} &
  \end{tabular}
  \end{center}
  \caption{The transverse momentum of the leading lepton at final selection in linear (a) and logarithmic (b) scale for the combination of $e^+e^-$, $\mu^+\mu^-$, and $e^{\pm}\mu^{\mp}$ channels.  The signal is shown for $m_H$=165~GeV and is scaled to the SM prediction for the combination of Higgs boson production from gluon fusion, vector boson fusion, and associated production.  The systematic uncertainty is shown after fitting.
  \label{fig:pt1}}
\end{figure*}

\begin{figure*}[htp]
  \begin{center}
  \begin{tabular}{lcr}
  \includegraphics[width=1.0\columnwidth]{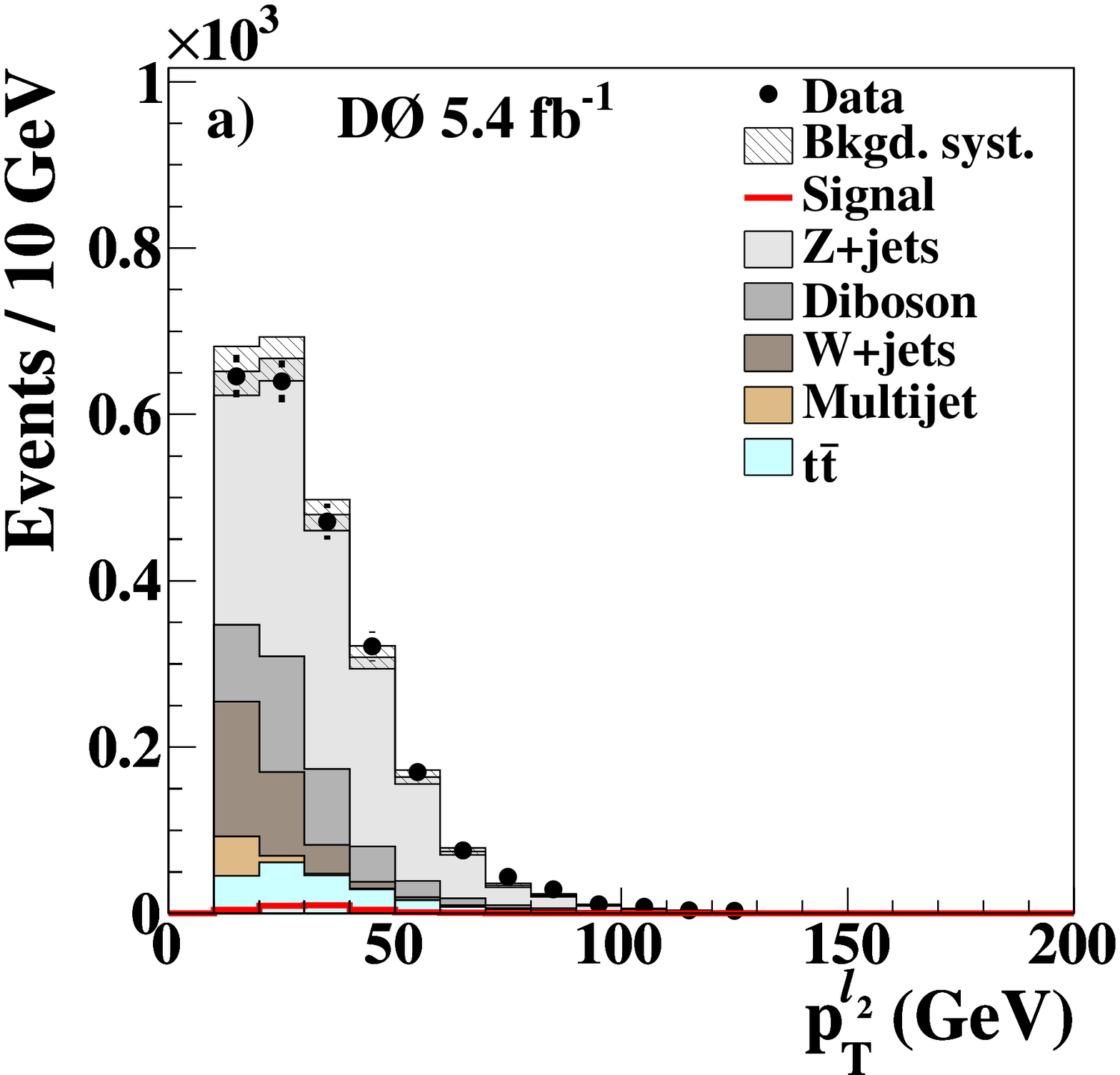} &
  \includegraphics[width=1.0\columnwidth]{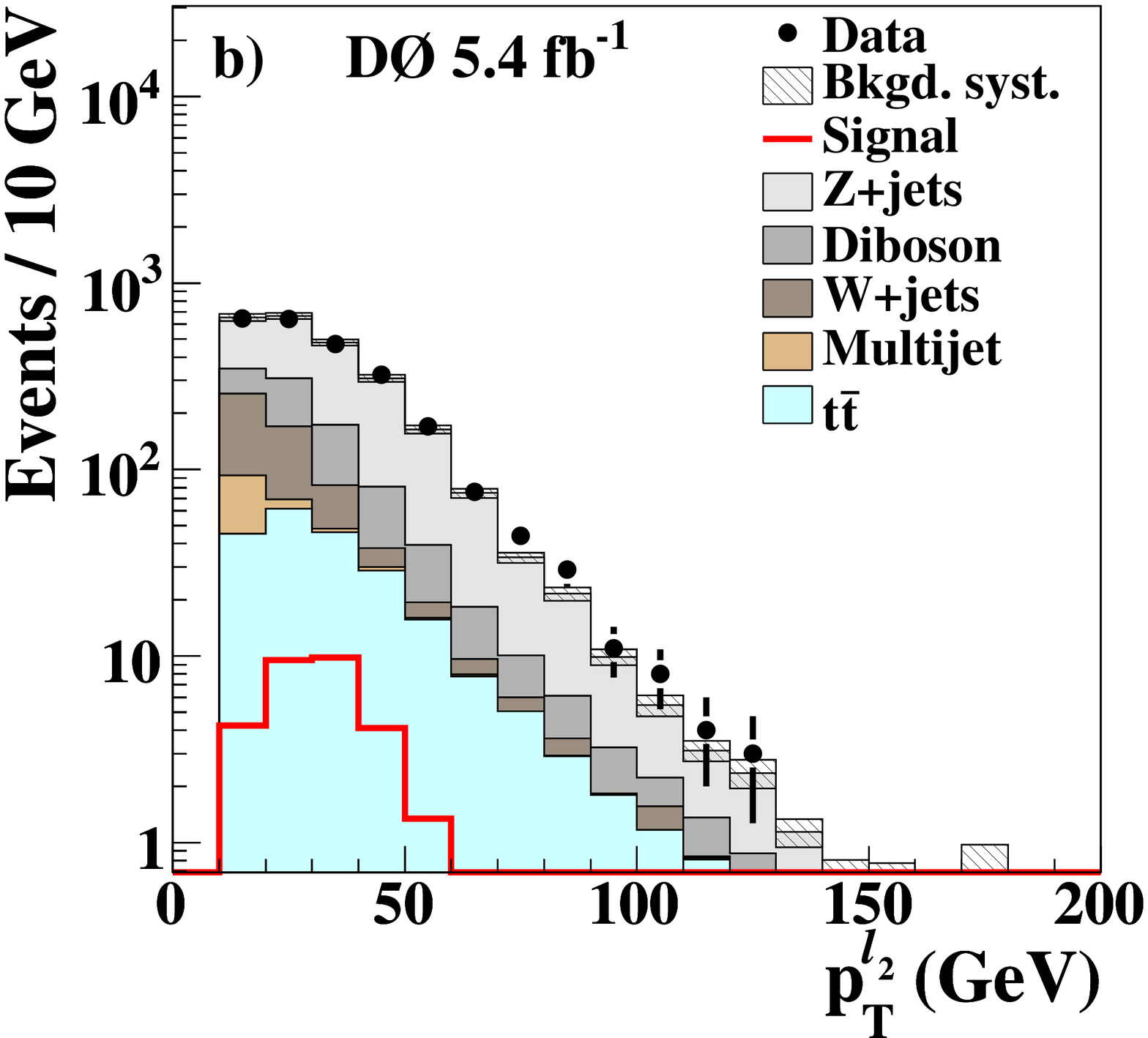} &
  \end{tabular}
  \end{center}
  \caption{The transverse momentum of the next-to-leading lepton at final selection in linear (a) and logarithmic (b) scale for the combination of $e^+e^-$, $\mu^+\mu^-$, and $e^{\pm}\mu^{\mp}$ channels.  The signal is shown for $m_H$=165~GeV and is scaled to the SM prediction for the combination of Higgs boson production from gluon fusion, vector boson fusion, and associated production.  The systematic uncertainty is shown after fitting.
  \label{fig:pt2}}
\end{figure*}

\begin{figure*}[htp]
  \begin{center}
  \begin{tabular}{lcr}
  \includegraphics[width=1.0\columnwidth]{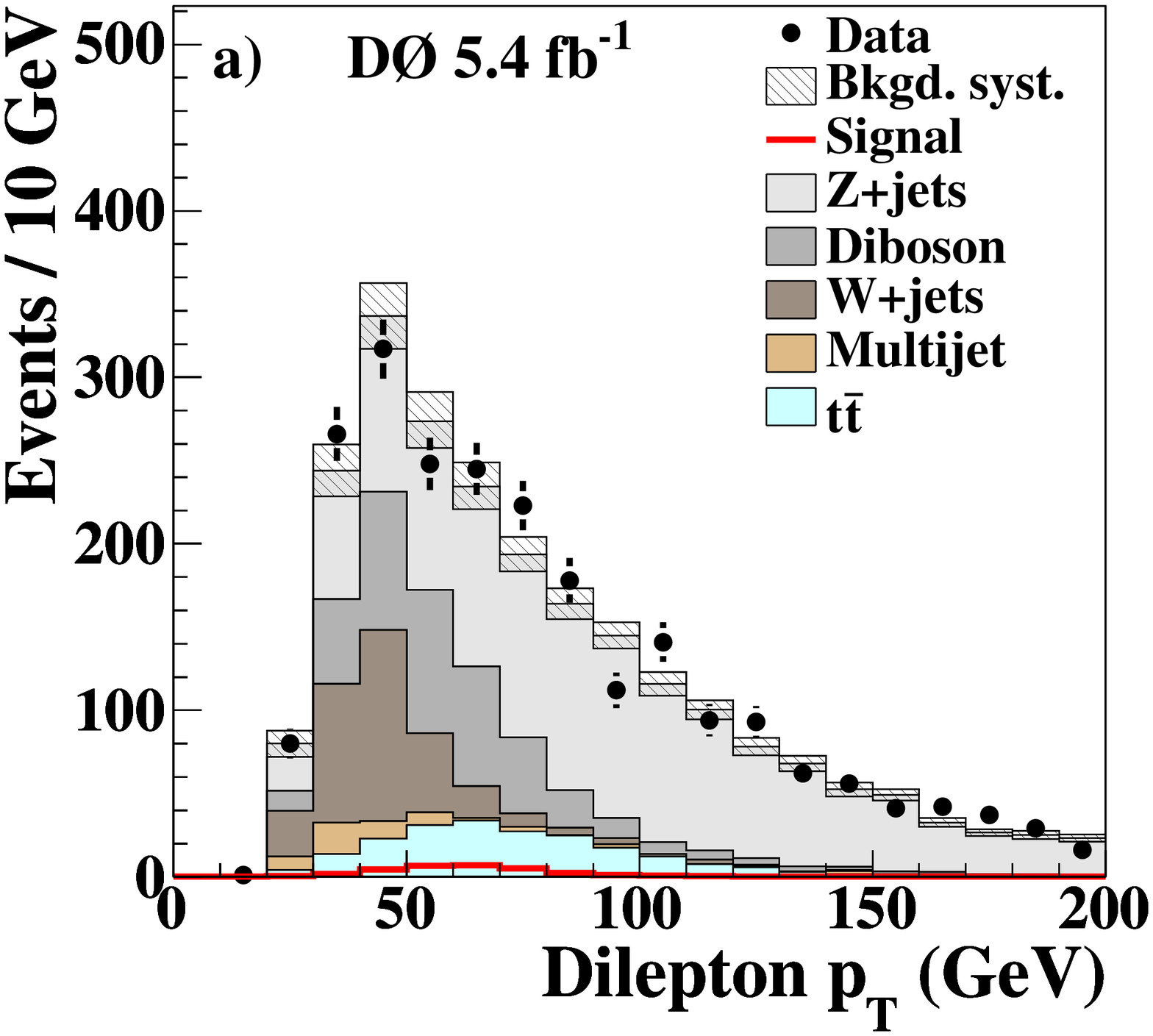} &
  \includegraphics[width=1.0\columnwidth]{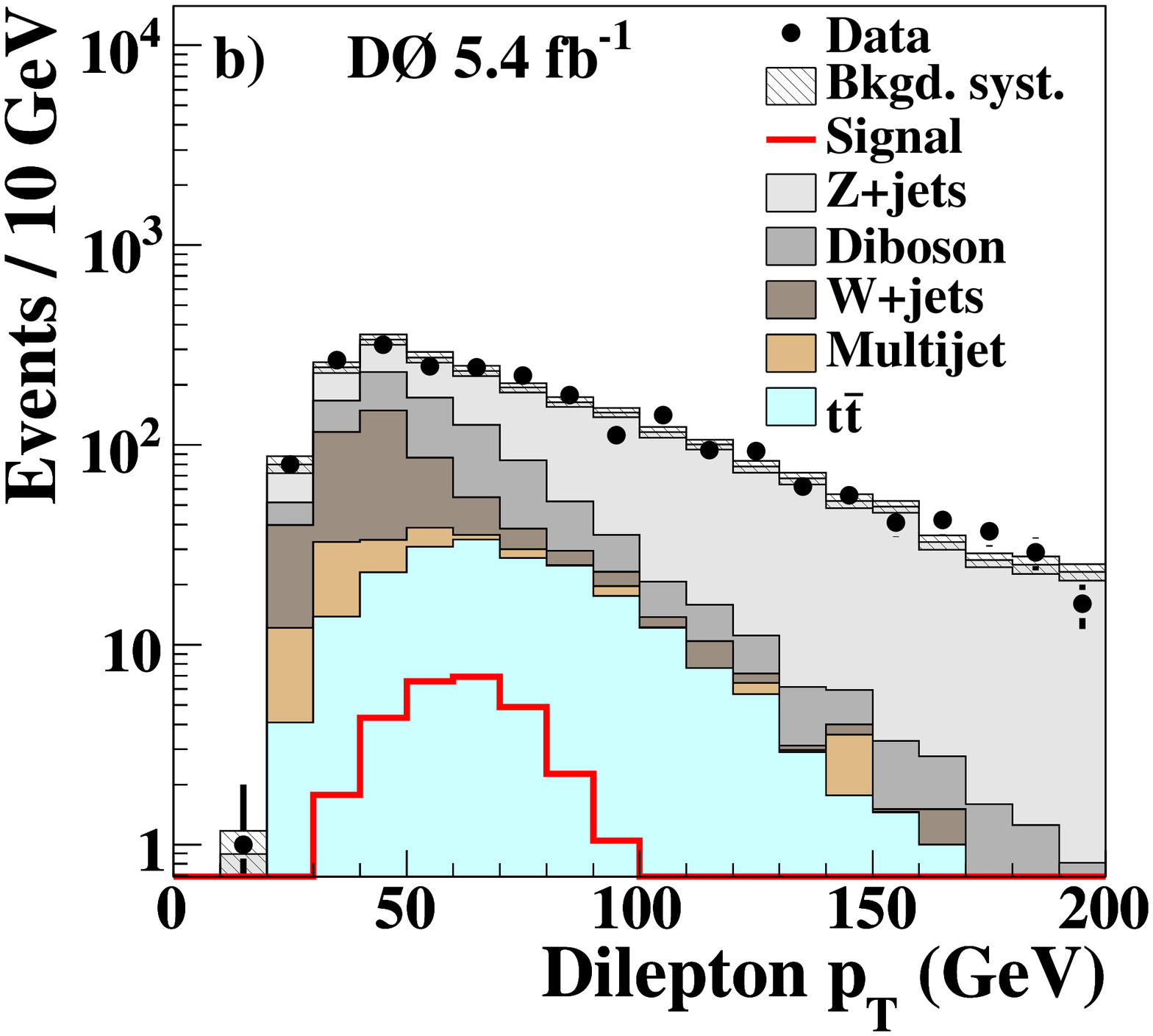} &
  \end{tabular}
  \end{center}
  \caption{The transverse momentum of the dilepton system at final selection in linear (a) and logarithmic (b) scale for the combination of $e^+e^-$, $\mu^+\mu^-$, and $e^{\pm}\mu^{\mp}$ channels.  The signal is shown for $m_H$=165~GeV and is scaled to the SM prediction for the combination of Higgs boson production from gluon fusion, vector boson fusion, and associated production.  The systematic uncertainty is shown after fitting.
  \label{fig:zpt}}
\end{figure*}

\begin{figure*}[htb]
 \begin{center}
 \begin{tabular}{lcr}
 \includegraphics[width=1.0\columnwidth]{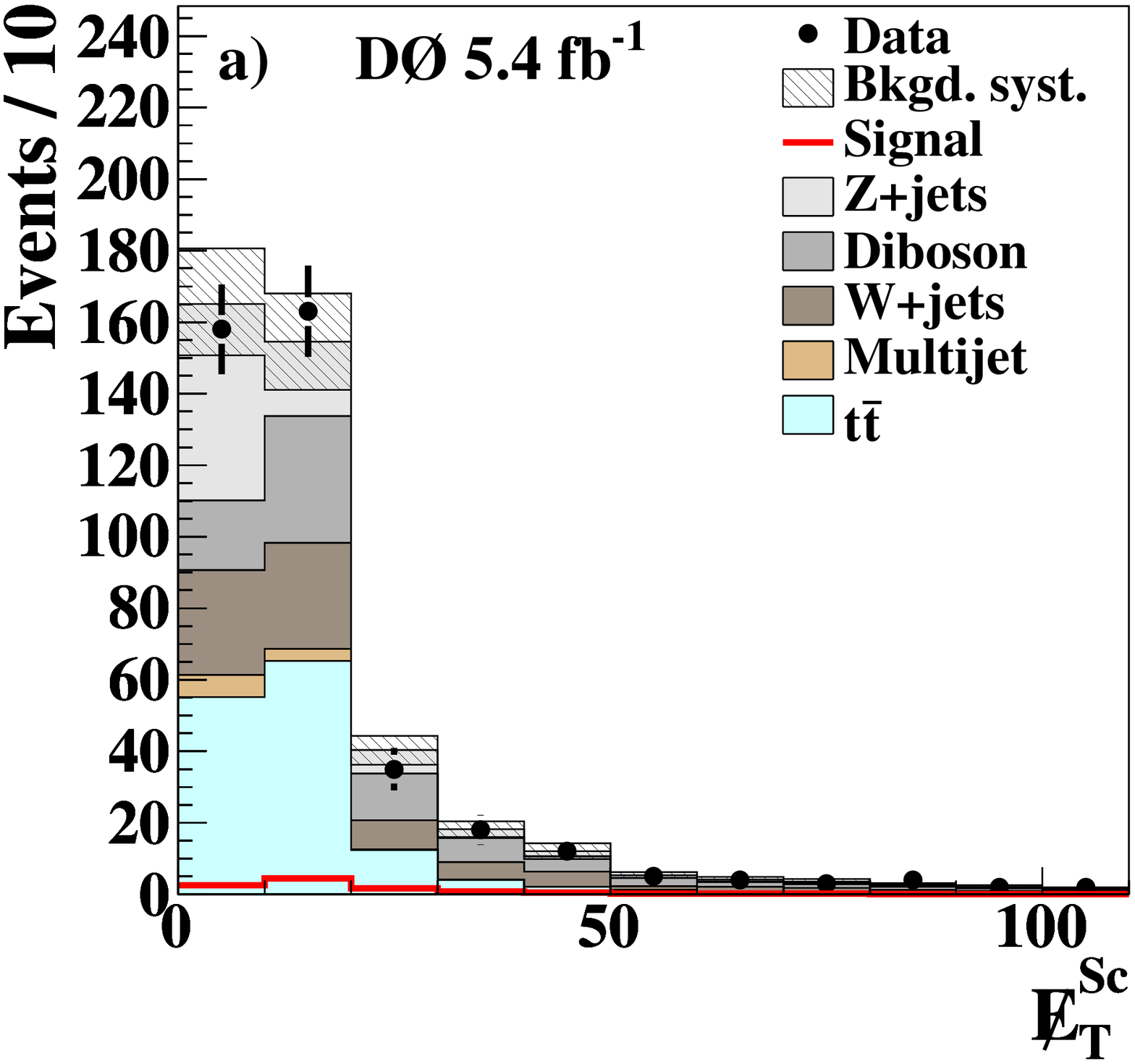} &
 \includegraphics[width=1.0\columnwidth]{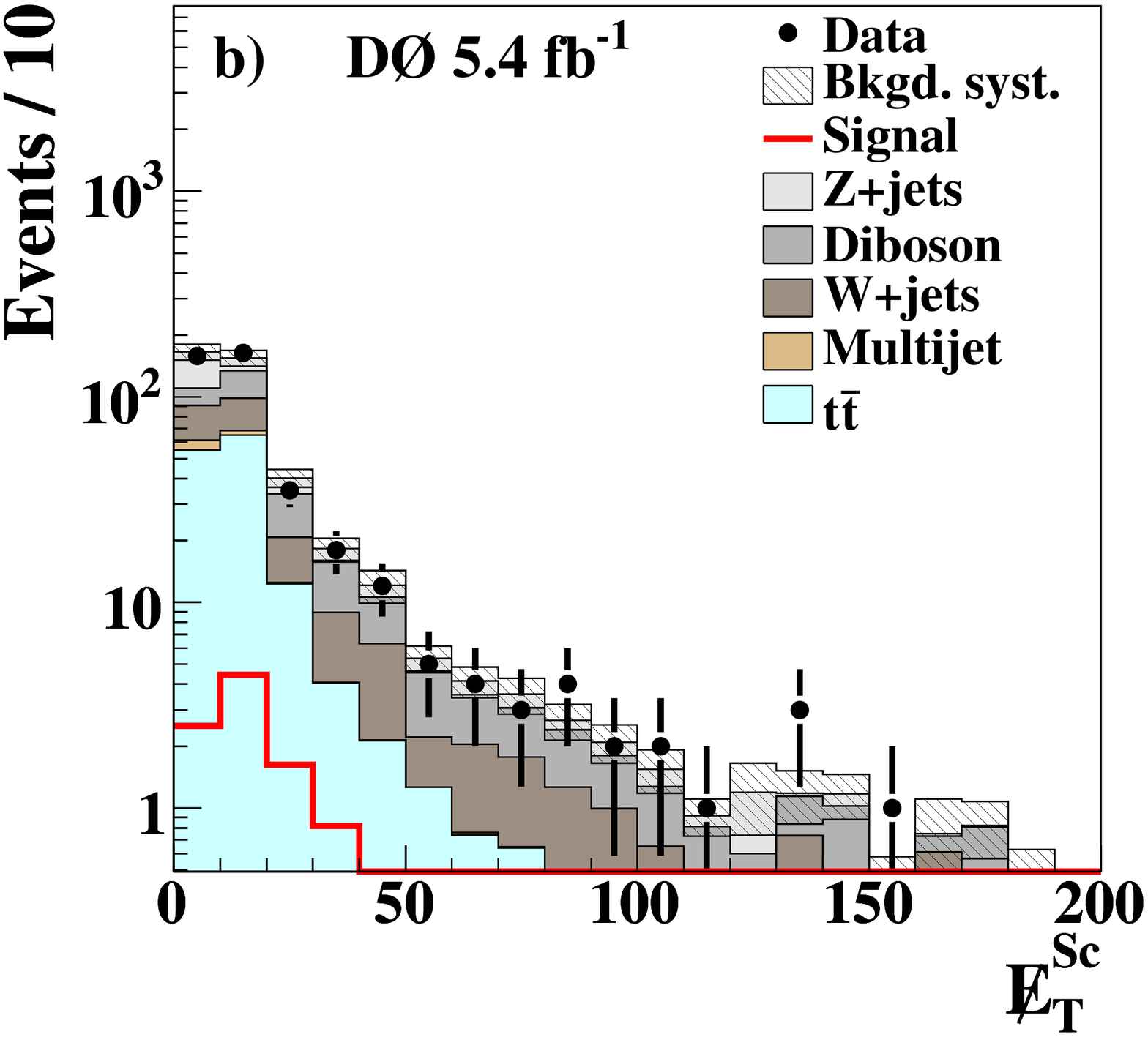} &
 \end{tabular}
 \end{center}
 \caption{The {\etmisssc} at final selection in linear (a) and
logarithmic (b) scale for the combination of $e^+e^-$
and $e^{\pm}\mu^{\mp}$ channels.  The signal is shown for
$m_H$=165~GeV and is scaled to the SM prediction for the combination
of Higgs boson production from gluon fusion, vector boson fusion, and
associated production.  The systematic uncertainty is shown after
fitting.
 \label{fig:metscal}}
\end{figure*}


\begin{thebibliography}{99}
\bibitem{d0coord}
The D0 coordinate system is cylindrical with the $z$ axis in the direction of the proton beam. 
In this frame, the angle $\phi$ is the azimuthal angle, and $\theta$ is the polar angle. The pseudorapidity $\eta$ 
is defined as  $\eta=-\ln[\tan(\theta/2)]$.
\end{thebibliography}
\end{document}